\documentclass[aps,prb,twocolumn,superscriptaddress,longbibliography,floatfix]{revtex4-2}

\usepackage[utf8]{inputenc}
\usepackage[T1]{fontenc}
\usepackage{amsmath,amssymb,bm}
\usepackage{graphicx}
\usepackage{xcolor}
\usepackage{xr-hyper}
\usepackage{hyperref}
\hypersetup{colorlinks=true,linkcolor=blue!55!black,citecolor=blue!55!black,urlcolor=blue!55!black,
  pdftitle={Momentum-selective graphene spectroscopy of moire excitons},
  pdfauthor={Katsunori Wakabayashi}}
\graphicspath{{../figures/}}
\newcommand{\qv}{\bm{q}}
\newcommand{\Rv}{\bm{R}}
\newcommand{\Gv}{\bm{G}}
\newcommand{\GM}{G_{\mathrm{M}}}
\newcommand{\aM}{a_{\mathrm{M}}}
\newcommand{\Pn}[1]{\bm{P}_{#1}}
\newcommand{\GG}[1]{\Gamma_{\mathrm{G},#1}}
\newcommand{\Gph}[1]{\Gamma_{\gamma,#1}}
\newcommand{\ii}{\mathrm{i}}
\newcommand{\dd}{\mathrm{d}}
\newcommand{\Reff}{R_{\mathrm{eff}}}

\begin{document}

\title{Momentum-selective graphene spectroscopy of moir\'e excitons}

\author{Katsunori Wakabayashi}
\affiliation{Research Center for Materials Nanoarchitectonics (MANA),
National Institute for Materials Science (NIMS), Namiki 1-1, Tsukuba 305-0044, Japan}
\email{WAKABAYASHI.Katsunori@nims.go.jp}

\date{\today}

\begin{abstract}
Far-field photoluminescence probes only the light-cone corner of an exciton's
momentum content, whereas energy transfer to nearby graphene integrates
in-plane momenta up to the inverse spacer thickness $1/z$.
For a WSe$_2$/MoSe$_2$ heterobilayer we build the transition polarization
$\bm{P}_n(\qv)$ of each localized moir\'e exciton from the registry-dependent
vector dipole texture and use each $\bm{P}_n$ for both baths.  The two
baths rank the states differently: the graphene-rate ratios differ from the
optical ones, their ordering inverts with spacer thickness, and the distance
exponents at $z=10$\,nm spread by more than two units.  These
ratios are independent of the transition-dipole prefactor.  Within the
$C_3$-symmetric leading-harmonic model the chiral $p_+$ and $d_-$ states are
also dark at normal incidence yet transfer to graphene at rates comparable to
the bright ground state.  Their graphene rate depends on the sign of the
texture phase, a quantity invisible in far-field emission, which we determine
from first principles.  In a device this prediction requires a splitting, larger than the
symmetry-breaking mixing, that protects the chiral partners.  Graphene proximity
thus acts as a momentum-selective near-field probe of moir\'e excitons, and we
propose its use as a practical spectroscopy.
\end{abstract}

\maketitle

\section{Introduction}
\label{sec:intro}

\begin{figure*}[tbp]
\includegraphics[width=\textwidth]{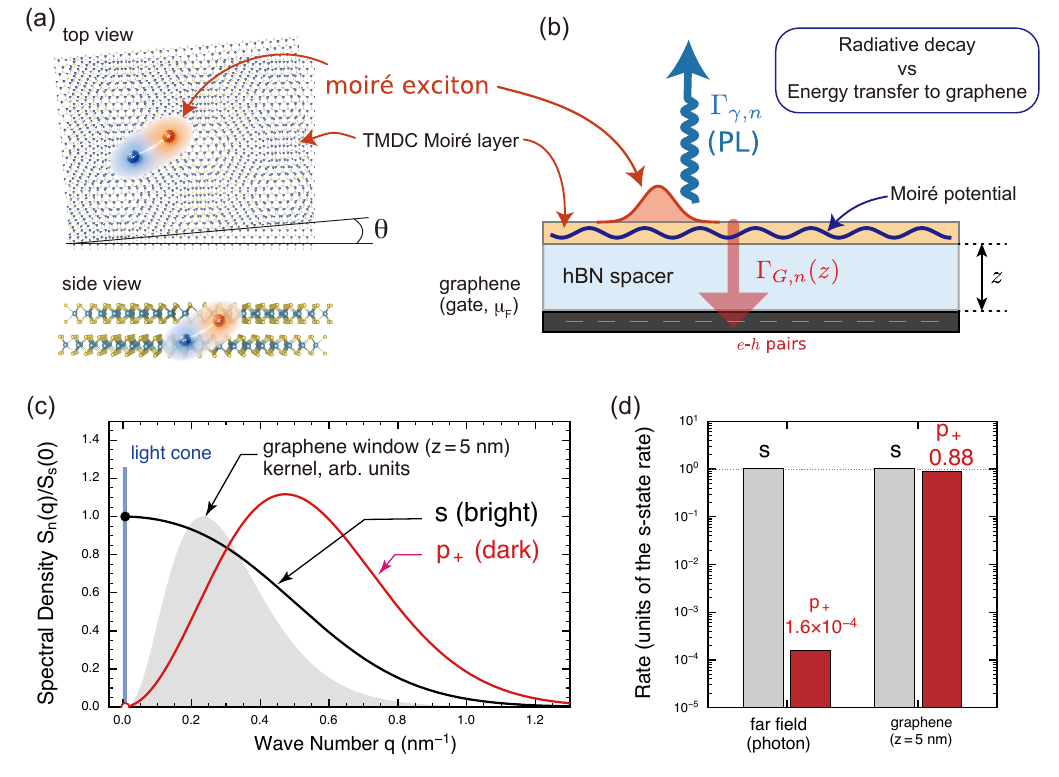}
\caption{Moir\'e exciton, device, and the momentum windows of the two baths.
(a)~Twisted WSe$_2$/MoSe$_2$ heterobilayer in top and side views ($\theta$
exaggerated) with an interlayer moir\'e exciton, whose electron and hole reside in
different layers.  (b)~Device geometry.  An hBN spacer of thickness $z$ separates
the exciton layer from a gated graphene sheet with Fermi level $\mu_F$.  The
exciton decays radiatively (photoluminescence, rate $\Gph{n}$) and by energy
transfer that creates electron--hole pairs in graphene (rate $\GG{n}(z)$).
Far-field photoluminescence and graphene energy
transfer probe complementary momentum sectors of one transition
polarization $\Pn{n}(\qv)$, and thus select different moir\'e-exciton
states.  (c)~The longitudinal spectral density
$S_n(q)=\langle|\hat{\qv}\cdot\Pn{n}(\qv)|^2\rangle_\varphi$ of the bright $s$
state and the chiral dark $p_+$ state (each normalized to $S_s(0)$).  Far-field
emission samples the light cone ($q\simeq0$, blue band; dots: values at $q\simeq0$),
where $S_s$ is finite but $S_{p_+}(0)=0$ by $C_3$ symmetry.  Graphene transfer integrates
$S_n(q)$ weighted by the near-field kernel $q^3e^{-2\gamma qz}$ (shaded, $z=5$\,nm,
normalized to its maximum for display), which captures the finite-$q$ weight of
\emph{both} states.  (d)~The resulting
rates, normalized to the $s$ state.  At
normal incidence $p_+$ is \emph{exactly}
dark within the $C_3$-symmetric leading-harmonic model.  Integrated over the
finite light cone, its radiative rate is $1.6\times10^{-4}$
of that of $s$ (the ``far-field'' bar), whereas its graphene rate remains
$0.88\,\GG{s}$.  Graphene thus reads a state that is strongly suppressed in
far-field emission.  The figure illustrates the momentum-window mechanism.  In a
device the darkness of the degenerate partner requires protection from $C_3$
breaking (Appendix~\ref{sm:strain}), whereas the nondegenerate $s$
benchmark is robust to it.}
\label{fig:concept}
\end{figure*}

Moir\'e superlattices of transition-metal-dichalcogenide (TMD) heterobilayers
host arrays of localized interlayer excitons [Fig.~\ref{fig:concept}(a)] whose energies, optical
selection rules, and dynamics are determined by the local atomic
registry~\cite{Rivera2015,Yu2017,Wu2018,Seyler2019,Tran2019,
RuizTijerina2020,MakShan2022}.  The moir\'e potential also reshapes intralayer and
hybridized excitons~\cite{Wu2017,Jin2019,Alexeev2019}, and these results build on the
well-characterized optics of monolayer TMDs~\cite{Wang2018,Matsuda2015,Xiao2012}.  We work in the low-density,
single-exciton regime, which excludes exciton--exciton annihilation~\cite{Mouri2014,Hoshi2017}.  Nearly all information on these states comes from far-field
optics, yet a photon of energy $\hbar\omega\simeq1.35$\,eV has an in-plane
escape-cone momentum of only $q_{\mathrm{light}}=\omega/c\simeq0.007$\,nm$^{-1}$,
two orders of magnitude below the moir\'e scale
$\GM=4\pi/(\sqrt{3}\aM)\simeq0.4$\,nm$^{-1}$, with $\aM$ the moir\'e period at a
twist angle $\theta=1^\circ$.  Thus, far-field
emission measures one value per state and polarization---the
$\qv\simeq0$ transition dipole---and is blind to the momentum texture that the
moir\'e potential and the registry-dependent optical matrix element imprint on
the exciton.

Nonradiative energy transfer~\cite{Forster1948} to a nearby graphene layer
[Fig.~\ref{fig:concept}(b)] probes a different window.  For a two-dimensional emitter at distance $z$ the
near field weights in-plane momenta as $q^3e^{-2\gamma qz}$ [$\gamma$ the
hexagonal-boron-nitride (hBN) anisotropy factor], peaking at
$q\simeq3/(2\gamma z)$.  A spacer of a few nanometers thus reads momenta of
$0.1$--$1$\,nm$^{-1}$, the moir\'e momentum scale.  Graphene quenching of
two-dimensional emitters is well established theoretically---the point-dipole
$z^{-4}$ law and the graphene-ruler concept~\cite{Swathi2008,Swathi2009,Chance1978,GomezSantos2011,Fezai2017}---and
experimentally, from individual-emitter quenching~\cite{Chen2010} and
distance-dependent transfer~\cite{Gaudreau2013,Tielrooij2015,Federspiel2015} to
gate-switchable energy flow in TMD/graphene
stacks~\cite{Froehlicher2018,Lorchat2020,Tebbe2024}.  Transfer from moir\'e-trapped
excitons to graphene has been analyzed within the random-phase approximation (RPA),
including dielectric-environment and distance effects, together with
TMD-emitter lifetime engineering near graphene~\cite{Hichri2021,Eddhib2021}.

Near-field transfer is known to be sensitive to finite-momentum dark
excitons~\cite{Malic2018} that far-field optics cannot reach~\cite{Katzer2023,Lin2023}, and
transfer from bright MoSe$_2$ excitons to graphene was recently found to vanish
already at $1$\,nm hBN spacing, with the residual coupling attributed to
finite-momentum hot excitons~\cite{Moghe2026}---evidence that the graphene
channel discriminates by momentum.  That experiment probes \emph{intralayer}
excitons at ${\sim}1$\,nm.  The registry-dependent \emph{vector} transition
polarization of the moir\'e states, and the state-resolved momentum selectivity
that follows from it, remain open.  Here we treat \emph{interlayer} moir\'e excitons at
$z\geq2$\,nm, where electrostatic coupling dominates Dexter and
charge-transfer processes, at the level of dimensionless ratios independent of
the exciton population.

In previous work~\cite{Wakabayashi2026quench} we developed a
minimal theory in which the finite exciton size enters through a
phenomenological Gaussian form factor $P_X(q)=D_XF_X(q)$.  The quenching rate
then measures a single transition-polarization radius $R_X$, and gating the
graphene Fermi level $\mu_F$ Pauli-blocks the transfer.  The present work
replaces the phenomenological form factor by the microscopic transition
polarization of each moir\'e exciton state of a WSe$_2$/MoSe$_2$ heterobilayer,
including states whose structure goes beyond a single radius.

The moir\'e
selection rules, the finite-momentum and dark-exciton sensitivity of graphene
transfer, and distance-dependent momentum filtering are each individually known.
Our specific contribution is to construct the finite-$q$ transition polarization
$\Pn{n}(\qv)$ of each localized moir\'e state from the registry-dependent
\emph{vector} dipole texture and to use each $\Pn{n}$ for both the photon
and the graphene bath.  Far-field photons and graphene then \emph{rank the
states differently}, because they sample complementary momentum windows of one
$\Pn{n}(\qv)$ [Fig.~\ref{fig:concept}(c)], turning graphene from a parasitic
quencher into a state- and momentum-selective probe.  Figure~\ref{fig:concept}(d)
shows the consequence for the $q=0$-dark $p_+$ state, whose far-field rate is
about $10^{-4}$ of the $s$ rate and whose graphene rate at $z=5$\,nm is $0.88$ of
the $s$ rate.  The transfer weights $\Pn{n}(\qv)$
by a smooth kernel whose width is fixed by the distance, and a distance series
constrains the momentum profile in this coarse-grained sense, without a unique
inversion.

We solve the exciton center-of-mass (COM) continuum model with the
\emph{ab initio} moir\'e potential and registry-dependent interlayer dipole of
Wu, Lovorn, and MacDonald (WLM)~\cite{Wu2018}, construct localized orbitals
$W_n(\Rv)$ of the lowest minibands, and evaluate
\begin{equation}
  \Pn{n}(\qv)=\sum_{j=1}^{3} \bm{d}_{j}\,\widetilde{W}_n(\qv+\bm q_j),
  \label{eq:Pq}
\end{equation}
where the corner-star amplitudes $\bm{d}_{j}$ of the local optical dipole
encode the registry and $\bm q_j$ are the three bright momenta at the $\kappa$
corners of the moir\'e Brillouin zone
(Appendix~\ref{app:derivation} derives Eq.~\eqref{eq:Pq} and the graphene
coupling from the interlayer optical matrix element).

Our central results are \emph{ratios} of rates, independent of the overall
transition-dipole scale.  Symmetry restricts the relative phase of the two
circular texture components, absent from the published first-principles fits,
to two values, and our first-principles calculation selects one of them.  (i)~The two baths rank the moir\'e states differently, and the
ordering of the graphene rates inverts with distance (Sec.~\ref{sec:selective}).
(ii)~Within the $C_3$-symmetric leading-harmonic model the optical selection
rules are exact at $q=0$.  The chiral $p_+$ and $d_-$ states are radiatively
suppressed yet couple to graphene at rates comparable to that of the bright $s$
state, provided that a protecting splitting preserves the chiral partners in a
device (Appendix~\ref{sm:strain}).  The
dark-partner graphene ratio changes fourfold with the sign of the texture phase.
Thus, graphene tests a property of the dipole texture that
far-field emission cannot access.

(iii)~The corner-star sampling determines the
far-field strengths (the $p_-$ oscillator strength is $3.9$ times that of $s$ at
$\theta=1^\circ$), and twist tunes every contrast (Sec.~\ref{sec:control}).
(iv)~The graphene-weighted spectral density defines a generalized
transition-polarization radius, distinct from the orbital size and undefined for
the dark states, through which bright states with positive low-$q$ curvature
recover the single-radius theory of Ref.~\onlinecite{Wakabayashi2026quench} at
large $z$ (Sec.~\ref{sec:validity-gauss}).

\section{Microscopic model}
\label{sec:model}

\subsection{Moir\'e center-of-mass model and localized states}
\label{sec:com}

We consider the interlayer exciton of a WSe$_2$/MoSe$_2$ heterobilayer at
small twist angle $\theta$, with moir\'e period $\aM\simeq a_0/\theta$
($a_0=0.332$\,nm is the monolayer lattice constant).  We use $\aM=18.9$\,nm ($\theta\simeq1.0^\circ$), referred to
as $\theta=1^\circ$ below.  The exciton center-of-mass Hamiltonian in the moir\'e
potential~\cite{Yu2017,Wu2018,Wakabayashi2026prr} is
\begin{align}
  H&=-\frac{\hbar^2\nabla_{\Rv}^2}{2M_X}+V_{\mathrm{M}}(\Rv),\nonumber\\
  V_{\mathrm{M}}(\Rv)&=2V_0\sum_{j=1}^{3}\cos(\Gv_j\!\cdot\!\Rv+\phi_V),
  \label{eq:H}
\end{align}
with $\Gv_j$ ($j=1,2,3$) the first-shell moir\'e reciprocal vectors at $120^\circ$
to one another, the exciton mass $M_X=m_e+m_h=0.84\,m_0$ (electron and hole
masses $m_e$ and $m_h$, free-electron mass $m_0$), and the first-harmonic potential derived from
density-functional theory,
$(V_0,\phi_V)=(11.8\,\mathrm{meV},79.5^\circ)$ for AA stacking
[$(1.8\,\mathrm{meV},155.2^\circ)$ for AB].  These parameters, the interlayer
dipole texture of Sec.~\ref{sec:texture}, and its conventions are taken from
Ref.~\onlinecite{Wu2018}.

AA and AB are the
near-$0^\circ$ (R-type) and near-$60^\circ$ (H-type) alignments, related by a
$\pi$ rotation of the top layer [Fig.~\ref{fig:bz}(a)].  The three
high-symmetry registries---metals aligned, Mo above the chalcogen, and Mo above
the hollow site---map onto $R_0=0$, $R_1=(\bm{a}_1+\bm{a}_2)/3$ (with $\bm a_1$, $\bm a_2$ the monolayer
lattice vectors of Appendix~\ref{sm:dft}), and
$R_2=2(\bm{a}_1+\bm{a}_2)/3$ [Fig.~\ref{fig:bz}(a,b)].  For AA stacking the potential minimum sits at
$R_1$ with depth $V_{\mathrm{M}}(R_1)=6V_0\cos(\phi_V+2\pi/3)=-66.7$\,meV.

\begin{figure*}[tbp]
\includegraphics[width=\textwidth]{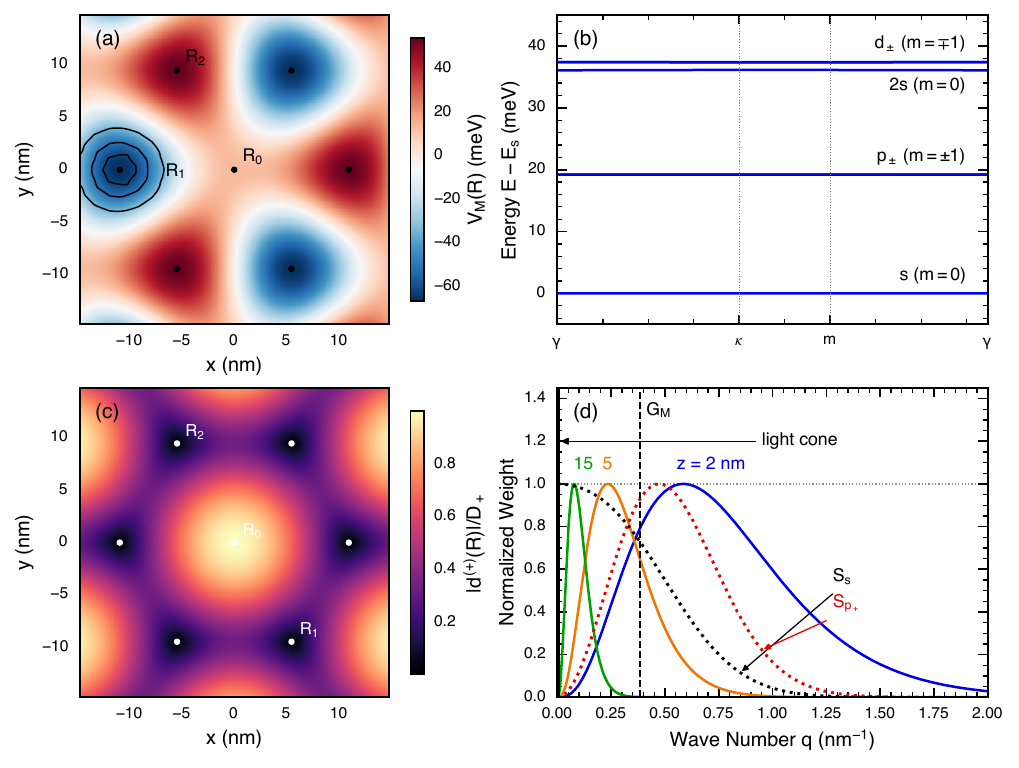}
\caption{Model overview for AA-stacked WSe$_2$/MoSe$_2$ at $\theta=1^\circ$.
(a)~Moir\'e potential $V_{\mathrm{M}}(\Rv)$ with the three registry sites and
the $s$ orbital $|W_s|$ (contours) at the potential minimum $R_1$.
(b)~Flat moir\'e minibands along $\gamma$--$\kappa$--$m$--$\gamma$ with their
$C_3$ quasi-angular-momentum labels.
(c)~Magnitude of the $\sigma^+$ dipole texture $d^{(+)}(\Rv)$
[Eq.~\eqref{eq:texture}].  The magnitude is maximal at $R_0$ and vanishes at $R_1$
and $R_2$.  Around the exciton site $R_1$ its phase winds once (a vortex; the phase
is not shown).
(d)~Momentum windows of the two baths.  The panel shows the graphene kernel $q^3e^{-2\gamma qz}$ for
three spacer thicknesses, the light cone, and the longitudinal spectral
densities $S_s(q)$ and $S_{p_+}(q)$ of the bright $s$ and the $q=0$-dark $p_+$
[Eq.~\eqref{eq:Sq}; dotted].  Each curve is normalized to its maximum, and the
vertical dashed line marks $\GM$.}
\label{fig:model}
\end{figure*}

Equation~\eqref{eq:H} is diagonalized in a plane-wave basis
(Appendix~\ref{app:numerics}).  At $\theta=1^\circ$ ($\aM=18.9$\,nm) the
lowest minibands are flat (widths $0.05$--$50\,\mu$eV) and are classified by
the $C_3$ quasi-angular momentum $m$ (defined modulo~3) about the exciton
site.  These are an $s$ band ($m=0$), a $p_\pm$ doublet ($m=\pm1$) at $19.2$\,meV, a
$2s$ band ($m=0$) at $36.1$\,meV, and a $d_\pm$ doublet ($m=\mp1$ classes of
the second shell) at $37.4$\,meV [Fig.~\ref{fig:model}(b)].  The point group
is $C_3$, whose irreducible representations are one dimensional, and the
degeneracy of the $m=\pm1$ partners follows from the complex-conjugation
symmetry of the real scalar COM Hamiltonian.  The complex conjugation acts as
an emergent spinless time reversal of the reduced single-valley model.
Physical time reversal connects the two valleys.

$C_3$-preserving perturbations that break this antiunitary symmetry split the
doublet.  The chiral COM states have the orbital angular momentum
$\langle L_z\rangle=\pm0.99$ about the trap center (gauge-free single-well solve;
Appendix~\ref{app:numerics}).  Thus, an out-of-plane magnetic field can split them.
However, the center of mass is charge neutral, and the splitting $\Delta_B$
arises from the constituent electron and hole moments, the
valley/Berry moment, and the relative motion.  We treat $\Delta_B$ as an external,
device-specific input (Secs.~\ref{sec:selective} and \ref{sec:protocol}).  The
level scheme agrees with the harmonic estimate $\hbar\Omega=21.1$\,meV and with
the localized-exciton picture of Refs.~\onlinecite{Wu2018,RuizTijerina2020}.

For each band (or degenerate pair) we construct unit-normalized localized
(Wannier) orbitals~\cite{Marzari2012} $W_n(\Rv)$ [Fig.~\ref{fig:model}(a)].  Because the bands are flat,
these are essentially the bound states of a single trap and are independent of
the projection gauge (Appendix~\ref{app:localization}).  The $s$-orbital size
$\ell_s\equiv\langle R^2\rangle^{1/2}=2.17$\,nm matches the root-mean-square localization length
$a_c\simeq2.2$--$2.3$\,nm reported by WLM~\cite{Wu2018}.  With this normalization a structureless dipole $\mu_0$ on a Gaussian orbital
of width $\ell$ gives $|\Pn{}(\qv)|^2=4\pi\ell^2\mu_0^2e^{-q^2\ell^2}$, the coherent
giant-oscillator-strength convention of
Refs.~\cite{Wakabayashi2026quench,Rashba1962,Feldmann1987,Andreani1991,Hanamura1988}.

\subsection{Registry-dependent dipole texture}
\label{sec:texture}

\begin{figure*}[tbp]
\includegraphics[width=\textwidth]{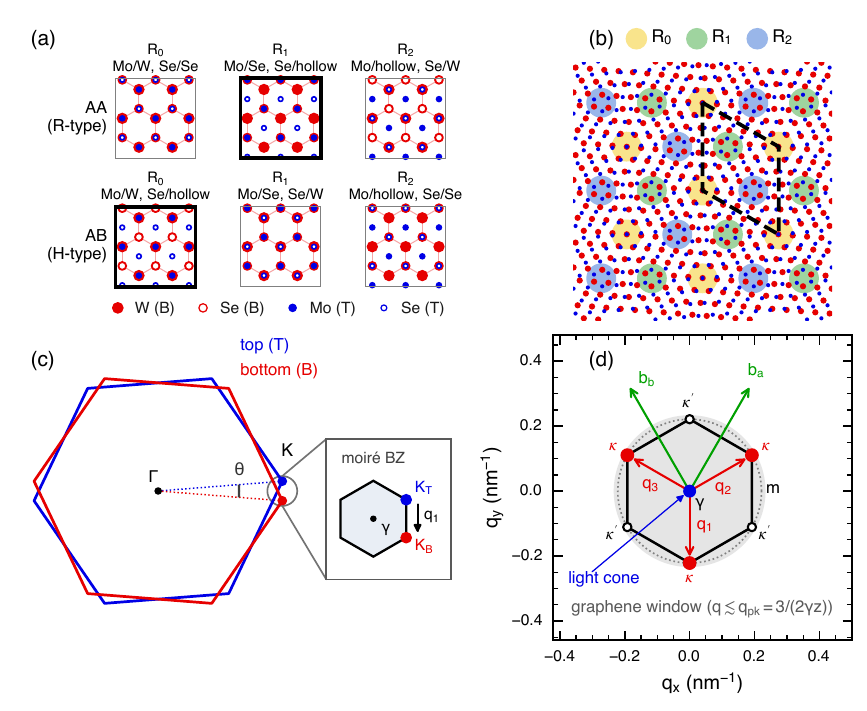}
\caption{Geometry of the twisted heterobilayer.
(a)~AA (R-type) and AB (H-type: top layer T rotated by $\pi$ about the metal
axis) stackings and the local arrangement at $R_n=n(\bm a_1+\bm a_2)/3$ (bold:
exciton site).  (b)~Twisted metal lattices ($\theta$ exaggerated), moir\'e cell
(dashed), and the regions of local displacement $R_0,R_1,R_2$ (yellow, green, and blue discs).  (c)~Monolayer Brillouin zones of the top (T) and bottom (B) layers, rotated by
$\pm\theta/2$ about $\Gamma$.  The two layer $K$ points (gray circle), split by
$\GM/\sqrt3$, are adjacent corners of the MBZ (framed, enlarged at right; center $\gamma$), and the vector
from $\bm K_T$ to $\bm K_B$ is $\bm q_1$ of panel (d).  The zone labels $\gamma$,
$\kappa$, and $m$ are unrelated to the anisotropy factor $\gamma$ and the quantum
number $m$.  (d)~The MBZ in the COM momentum $\bm q$.  The red arrows are the $C_3$-closed
star of bright momenta $\bm q_1=\bm K_B-\bm K_T$ (along $-q_y$), $\bm q_2$, and $\bm q_3$
($+120^\circ,+240^\circ$), pointing from $\gamma$ to the $\kappa$ corners (filled;
$|\bm q_j|=\GM/\sqrt3=K\theta$; dotted circle: $|\bm q|=K\theta$).  The open circles are the $\kappa'$ corners, where the
bright excitons of the $-K$ valley lie~\cite{Wu2018}, and the green arrows are the
reciprocal vectors $\bm b_a=\bm q_2-\bm q_1$ and $\bm b_b=\bm q_3-\bm q_1$ ($|\bm b|=\GM$).
Far-field emission samples only the light cone (blue disk at $\gamma$).  Graphene reads the
shaded disk $q\lesssim q_{\mathrm{pk}}=3/(2\gamma z)$, the peak of the near-field kernel
($z=5$\,nm), which reaches the corners.}
\label{fig:bz}
\end{figure*}

The momentum-space setup is summarized in Fig.~\ref{fig:bz}(c,d).  The interlayer
optical matrix element varies with the local registry, and the bright
momenta of the interlayer exciton are the three $\kappa$ corners of the
moir\'e Brillouin zone (MBZ), $\bm q_1,\bm q_2,\bm q_3$
($|\bm q_j|=K\theta=\GM/\sqrt3$, with $K=4\pi/(3a_0)$ the monolayer
Dirac-point momentum), a $C_3$-closed star whose centroid
coincides with the minimum of the COM kinetic energy
(Appendix~\ref{app:derivation}).  In the frame of
Eq.~\eqref{eq:H}, where the kinetic term is $\hbar^2q^2/2M_X$, the
leading-harmonic texture consists, for each circular
polarization $\sigma^\pm$, of three Fourier components of equal weight
\emph{on the corner star}:
\begin{equation}
  d^{(\pm)}(\Rv)=\frac{D_\pm}{3}\,e^{\ii\bm{c}\cdot\Rv}
  \Bigl[1+e^{-\ii\bm{b}_a\cdot(\Rv-R^{\pm}_c)}
        +e^{-\ii\bm{b}_b\cdot(\Rv-R^{\pm}_c)}\Bigr],
  \label{eq:texture}
\end{equation}
with harmonics at $\{\bm c,\bm c-\bm b_a,\bm c-\bm b_b\}=
\{-\bm q_1,-\bm q_2,-\bm q_3\}$.  Here $\bm{b}_a,\bm{b}_b$ are adjacent
first-shell moir\'e vectors ($60^\circ$ apart),
$\bm c=(\bm b_a+\bm b_b)/3=-\bm q_1$ is the star centroid of the
undressed harmonics, and the channel centers are the registries at which
the respective polarization is locally allowed.  These centers are $R^{+}_c=R_0$ and
$R^{-}_c=R_1$, with $(|D_+|,|D_-|)=(0.22,0.06)\,\mathcal{D}'$ for AA
stacking ($\mathcal{D}'$ is the MoSe$_2$ intralayer transition-dipole
scale).

The magnitude $|d^{(\pm)}(\Rv)|$ is invariant under the plane-wave dressing
$e^{\ii\bm c\cdot\Rv}$.  Each channel still vanishes at the two
high-symmetry sites other than its own center, reproducing the
local selection rules $\bm d(R_0)\propto D_+\bm e_+$,
$\bm d(R_1)\propto D_-\bm e_-$,
$\bm d(R_2)=0$~\cite{Yu2017,Wu2018}
[$\bm e_\pm=(\hat x\pm\ii\hat y)/\sqrt2$ is the circular basis].
The dressing is nevertheless essential for the momentum
bookkeeping.  The undressed harmonics
$\{0,\bm b_a,\bm b_b\}$ require the shifted kinetic term
$\hbar^2|\qv+\bm q_1|^2/2M_X$, and combining them with the
unshifted kinetic term breaks the $C_3$ covariance of the coupling
(Appendix~\ref{app:derivation}).  With Eq.~\eqref{eq:texture} the vector field
$\bm d(\Rv)=d^{(+)}\bm e_++d^{(-)}\bm e_-$ is exactly covariant under the geometric
$120^\circ$ rotation about every high-symmetry site (verified to $10^{-13}$,
Appendix~\ref{app:numerics}), and the exact optical selection rules
follow.

The published first-principles fits give the magnitudes $(|D_+|,|D_-|)$.  The relative phase
$\beta=\arg(D_-/D_+)$ is restricted by the valley-preserving antiunitary
symmetry of the R-stacked bilayer (time reversal combined with a vertical
mirror) to $\beta\in\{0,\pi\}$.  A first-principles calculation of the
aligned bilayer gives $\beta=0$ and reproduces the published magnitudes $(|D_+|,|D_-|)$ and the
moir\'e potential to within a few percent (Appendix~\ref{sm:dft}).  We use
$\beta=0$ in all results below, and the appendices and the Supplemental
Material~\cite{SM} give the $\beta=\pi$ values as a sensitivity check.

For AA stacking the exciton sits at $R_1$.  There the $\sigma^-$ texture is
\emph{aligned} (no phase winding), whereas the $\sigma^+$ texture has a
first-order zero with winding number one---a vortex
[Fig.~\ref{fig:model}(c)].  Because the dressed texture is $C_3$ covariant
about $R_1$, each channel belongs to a \emph{pure} local angular-momentum
class (mod~3) there.  At $\qv=0$ the $\sigma^-$ channel couples only to $m=0$
states and the $\sigma^+$ channel only to $m=-1$ states (Sec.~\ref{sec:rules}).

The transition polarization of state $n$ is the in-plane vector field
$\Pn{n}(\qv)=P^{(+)}_n(\qv)\bm e_+ + P^{(-)}_n(\qv)\bm e_-$ with the channel
amplitudes given by Eq.~\eqref{eq:Pq}, evaluated from the Bloch coefficients
through the exact identity
$\widetilde W_n(\bm k+\Gv)=\sqrt{A_c}\,\widetilde C_{n\bm k}(\Gv)$, with $A_c$ the
moir\'e cell area and $\widetilde C_{n\bm k}(\Gv)$ the plane-wave coefficients of the
Bloch states (Appendix~\ref{app:numerics}).  The out-of-plane component of the interlayer \emph{transition} dipole is
small for the $K$-valley $\sigma^\pm$ transitions considered here (the
large static dipole of the interlayer exciton is a permanent dipole, distinct
from the transition dipole), and we work with the in-plane
components throughout.

\subsection{Photon and graphene kernels}
\label{sec:kernels}

Both decay channels are functionals of one $\Pn{n}(\qv)$.  The far-field
radiative rate samples the light cone, which is pointlike on every structural
scale of $\Pn{n}$, and the two circular polarizations feed orthogonal photon
modes.  Thus, per channel
\begin{equation}
  \Gph{n}^{(\pm)}\;\propto\;f_n^{(\pm,0)}\equiv\bigl|P_n^{(\pm)}(0)\bigr|^{2},
  \label{eq:photon}
\end{equation}
where $f_n^{(\pm,0)}$ is the normal-incidence oscillator strength per
circular channel (the light-cone-integrated rate is taken up in
Sec.~\ref{sec:rules}).
Resolving the individual channels experimentally corresponds to
polarization-selective detection near normal emission.  Graphene, by
contrast, couples electrostatically to the \emph{transition
charge}.  Continuity relates the interband transition-current density to the
transition polarization, $\rho_X(\qv)=-\ii\qv\cdot\Pn{n}(\qv)$
(Appendix~\ref{app:derivation}).  Thus, only the longitudinal projection
$\hat q\cdot\Pn{n}$ drives the transfer:
\begin{equation}
  \GG{n}(z)=C_{\mathrm{G}}(\omega)\int^{q<q_c}
  \frac{\dd^2q}{(2\pi)^2}\;q^{2}e^{-2\gamma qz}\,
  \bigl|\hat q\cdot\Pn{n}(\qv)\bigr|^{2}.
  \label{eq:graphene}
\end{equation}
Here $C_{\mathrm{G}}(\omega)$ collects the Coulomb and graphene-response
prefactors (Appendix~\ref{app:derivation}).

With $\hat q=\qv/q$ and $\hat q\cdot\bm e_\pm=e^{\pm\ii\varphi_q}/\sqrt2$
($\varphi_q$ the polar angle of $\qv$),
\begin{align}
  \bigl|\hat q\cdot\Pn{n}\bigr|^{2}
  =\tfrac12\Bigl[&|P^{(+)}_n|^2+|P^{(-)}_n|^2\nonumber\\
  &+2\,\mathrm{Re}\bigl(P^{(+)}_nP^{(-)*}_ne^{2\ii\varphi_q}\bigr)\Bigr].
  \label{eq:longitudinal}
\end{align}
The interchannel interference term survives the angular integration
through the harmonic $l=-2$ of $P^{(+)}_nP^{(-)*}_n$.  About the trap center the
$\sigma^+$ and $\sigma^-$ amplitudes of a state of quasi-angular momentum $m$
belong to the classes $l\equiv m+1$ and $l\equiv m$ (mod~3), and their product
lies in the class $l\equiv+1$, which contains $l=-2$.  The interference is
allowed for every state and depends on $\beta$.  Equivalently, the
longitudinal amplitude $\hat q\cdot\Pn{n}$ belongs to a single $C_3$ class, as
the transition charge of a $C_3$ eigenstate must.

The $C_3$ classification also makes the photon and graphene decay matrices of the
degenerate $p$ doublet diagonal in the chiral basis (Sec.~\ref{sec:selective}).

The kinematic cutoff is $q_c=\omega/v_F$, with $v_F$ the graphene Fermi velocity~\cite{CastroNeto2009}
($\hbar v_F=0.658$\,eV\,nm,
$q_c=2.05$\,nm$^{-1}$ at $\hbar\omega=1.35$\,eV).  The hBN anisotropy factor is
$\gamma=\sqrt{\varepsilon_\parallel/\varepsilon_\perp}=1.28$, with the in-plane and
out-of-plane optical permittivities $\varepsilon_\parallel=4.98$ and
$\varepsilon_\perp=3.03$~\cite{Laturia2018,Wakabayashi2026quench}.  We include
$\gamma$ in every rate, and all distances refer to one physical spacer thickness
$z$.  The neutral interband kernel $q^2$ is upgraded in
Sec.~\ref{sec:validity-gauss} to the Dirac-RPA loss function of doped
graphene~\cite{Ando2006,Wunsch2006,Hwang2007},
$q^2\to k_F\,L(q/k_F,\hbar\omega/E_F)$ with $E_F=\hbar v_Fk_F=|\mu_F|$, which
adds Pauli blocking~\cite{Falkovsky2007}.  All rates use the state-resolved
transition energy $\hbar\omega_n=\hbar\omega_s+(E_n-E_s)$
($\hbar\omega_s=1.35$\,eV).  The radiative rate is proportional to $\omega_n^{3}$,
$C_{\mathrm{G}}\propto1/\omega_n$ [from $-\mathrm{Im}\,\chi=q^{2}/(4\hbar\omega)$, with $\chi$ the graphene density
response], and $q_{c,n}=\omega_n/v_F$.  The normal-incidence oscillator strengths
$f^{(0)}_n\equiv\sum_\pm|P^{(\pm)}_n(0)|^2$ are independent of $\omega_n$.

The scales $C_{\mathrm{G}}$ and $\mathcal{D}'$ drop out of every ratio quoted
below---state contrasts $\GG{n}/\GG{s}$, the local slope
$\alpha_n(z)=-\dd\ln\GG{n}/\dd\ln z$, and the normal-incidence
oscillator-strength selectivity
\begin{equation}
  B^{(0)}_n(z)=\frac{\GG{n}(z)/f^{(0)}_n}{\GG{s}(z)/f^{(0)}_s},
  \label{eq:Bn}
\end{equation}
with $f^{(\pm,0)}_n=|P^{(\pm)}_n(0)|^2$ the two circular-channel parts of $f^{(0)}_n$.  In any theory in which the states
share a common form factor and differ only in dipole magnitude---the
phenomenology of Ref.~\onlinecite{Wakabayashi2026quench}---$B^{(0)}_n\equiv1$.
Thus, $B^{(0)}_n-1$ directly measures the state-resolved momentum structure.
The selectivity $B^{(0)}_n\to1$ for every bright state at large $z$, where the kernel collapses
onto $q=0$.  Replacing $f^{(0)}_n$ by the light-cone-integrated rate gives the
physical selectivity $B^{\mathrm{LC}}_n$, finite for every state including the
$q=0$-dark ones [Fig.~\ref{fig:select}(b)].  The scales $C_{\mathrm{G}}$ and $\mathcal{D}'$ recombine into
known constants in the absolute branching ratio $Q_n(z)=\GG{n}/\Gph{n}$ of the
bright states and in $\GG{n}/\Gph{s}$ for the dark ones (Sec.~\ref{sec:protocol}).

\section{Selection rules of the textured states}
\label{sec:rules}

The convolution structure of Eq.~\eqref{eq:Pq} makes the optical selection
rules explicit.  At $\qv=0$,
$P_n^{(\pm)}(0)=\sum_{j} d^{(\pm)}_{j}\widetilde W_n(\bm q_j)$ samples the
envelope at the three corner momenta $\bm q_j$, all at radius
$|\bm c|=\GM/\sqrt3$ from the momentum center of the trapped state.
Because the dressed texture is $C_3$ covariant about the trap, the three
samples add with pure $C_3$ phases and the selection rules are
\emph{exact}---in the sense of holding identically within the
$C_3$-symmetric leading-harmonic model, for any radial shape and texture
magnitudes (the effect of symmetry-breaking perturbations is quantified in
Appendix~\ref{sm:strain}).  They read
\begin{equation}
  \sigma^-\!\leftrightarrow m=0\;(s,2s),\quad
  \sigma^+\!\leftrightarrow m=-1\;(p_-,d_+),
  \label{eq:smearing}
\end{equation}
while the $m=+1$ states ($p_+$, $d_-$) are \emph{exactly dark in both
channels at $\qv=0$}.  These rules reproduce the classification of Wu, Lovorn,
and MacDonald~\cite{Wu2018}---each moir\'e
absorption peak appears in a single circular polarization, and one member
of each corner triad is optically forbidden---and we
verify them numerically to better than $10^{-8}$ of the $s$-state strength
(Table~\ref{tab:states}; Appendix~\ref{app:numerics}).

The corner sampling also determines the bright-state \emph{strengths}.  For the
$s$ orbital the samples lie at radius $\GM/\sqrt3$, whereas for $p_-$ they lie
near the finite-momentum maximum of the $p$ envelope.  At $\theta=1^\circ$ the
$p_-$ line has $3.9$ times the $s$ oscillator strength ($4.1$ times
its radiative rate).  The $2s$ and $d_+$ oscillator strengths
are $0.71$ and $1.15$ of the $s$ value
(Table~\ref{tab:states}).

Because helicity is locked to the $C_3$ class, polarization-resolved emission
sorts the bright manifold ($\sigma^-$ for $s,2s$ and $\sigma^+$ for
$p_-,d_+$).  Graphene also reaches the $q=0$-dark $p_+,d_-$ partners.
Thus, polarization and finite-$q$ transfer provide complementary labels
for the six states [Fig.~\ref{fig:select}(a)].

Inside the finite light cone the dark-state amplitudes grow again and give a
small radiative tail.  Integration over the ordinary and extraordinary photon cones in homogeneous
hBN gives
$\Gamma^{\mathrm{LC}}_{\gamma,p_+}=1.6\times10^{-4}\,\Gph{s}$
and $\Gamma^{\mathrm{LC}}_{\gamma,d_-}=8.1\times10^{-3}\,\Gph{s}$
(Appendix~\ref{app:numerics}).

For a degenerate doublet, generic $C_3$ breaking mixes the chiral partners.
Observing the dark-partner contrast then requires a prior
$C_3$-preserving splitting larger than the mixing
(Appendix~\ref{sm:strain}).

\begin{table*}[tbp]
\caption{State-resolved observables for AA-stacked WSe$_2$/MoSe$_2$ at
$\theta=1^\circ$ ($\aM=18.9$\,nm), with the longitudinal kernel
[Eqs.~\eqref{eq:graphene} and \eqref{eq:longitudinal}].  $m$ is the $C_3$
quasi-angular momentum (mod~3).  Normal-incidence oscillator strengths are quoted per circular
channel, normalized to the total $s$-state oscillator strength
$f_s^{(0)}=f_s^{(+,0)}+f_s^{(-,0)}$.  Graphene rates at $z=5$\,nm are
normalized to the $s$ state.  The quantity $B^{(0)}_n$ is the oscillator-strength-normalized
selectivity [Eq.~\eqref{eq:Bn}], and $\alpha_n$ is the local distance exponent at
$z=10$\,nm.  All rates use the state-resolved transition energies
$\hbar\omega_n$ (Sec.~\ref{sec:kernels}).  Photon columns are the normal-incidence oscillator strengths
$f_n^{(0)}$ per circular channel.  Entries ``0'' are exact at $\qv=0$ within
the $C_3$-symmetric leading-harmonic model ($C_3$ selection, Eq.~\eqref{eq:smearing}; numerical residuals below
$10^{-8}$), with the light-cone radiative tails of the $q=0$-dark $p_+$
and $d_-$ given in Sec.~\ref{sec:rules}.  For these states the
normal-incidence $B^{(0)}_n$ is formally infinite ($f_n^{(0)}=0$).  The
value in parentheses is the finite light-cone selectivity $B^{\mathrm{LC}}_n$
that normalizes by the light-cone-integrated radiative tail
($5.5\times10^{3}$ for $p_+$ and $2.3\times10^{2}$ for $d_-$,
Sec.~\ref{sec:selective}).  The absolute graphene scale is quoted in
Sec.~\ref{sec:protocol}.}
\label{tab:states}
\begin{ruledtabular}
\begin{tabular}{lccccccc}
$n$ & $E_n\!-\!E_s$ & $m$ &
$\dfrac{f_n^{(+,0)}}{f_s^{(0)}}$ & $\dfrac{f_n^{(-,0)}}{f_s^{(0)}}$ &
$\dfrac{\GG{n}}{\GG{s}}$ & $B^{(0)}_n$ & $\alpha_n$ \\
 & (meV) & & & & ($z\!=\!5$) & ($z\!=\!5$) & ($z\!=\!10$) \\
\hline
$s$   & $0$    & $0$  & $0$    & $1.00$ & $1.00$ & $1.0$  & $3.9$ \\
$p_+$ & $19.2$ & $+1$ & $0$    & $0$    & $0.88$ & $\infty$\,($5.5{\times}10^{3}$) & $5.5$ \\
$p_-$ & $19.2$ & $-1$ & $3.92$ & $0$    & $2.18$ & $0.55$ & $3.5$ \\
$2s$  & $36.1$ & $0$  & $0$    & $0.71$ & $2.00$ & $2.8$  & $4.5$ \\
$d_+$ & $37.4$ & $-1$ & $1.15$ & $0$    & $0.48$ & $0.42$ & $3.0$ \\
$d_-$ & $37.4$ & $+1$ & $0$    & $0$    & $1.89$ & $\infty$\,($2.3{\times}10^{2}$) & $5.1$ \\
\end{tabular}
\end{ruledtabular}
\end{table*}

\section{State-selective quenching}
\label{sec:selective}

\begin{figure*}[tbp]
\includegraphics[width=\textwidth]{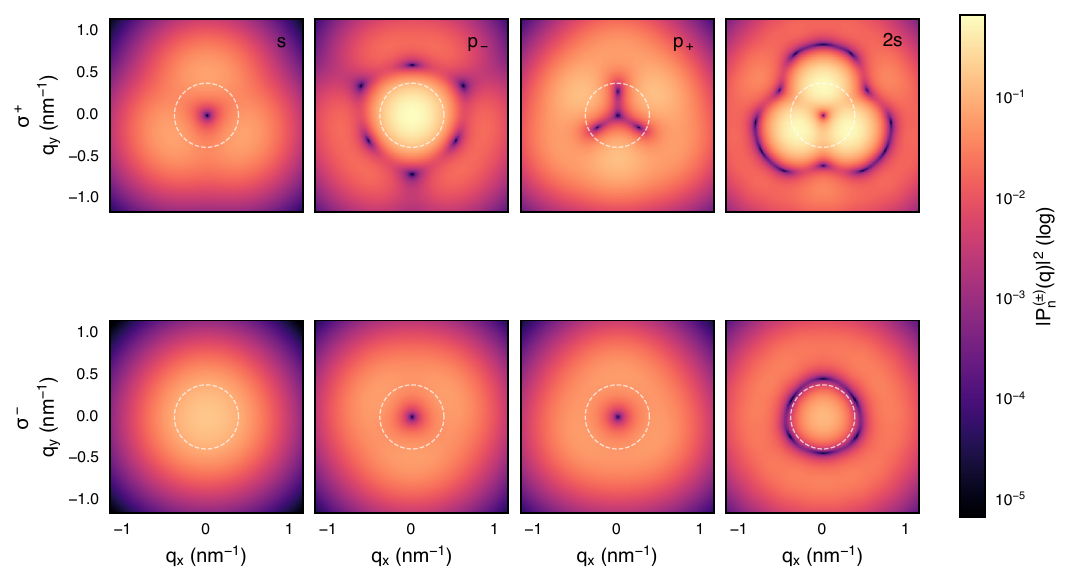}
\caption{Microscopic transition polarization $|P_n^{(\pm)}(\qv)|^2$
(logarithmic color scale) of the four lowest moir\'e states for AA stacking
at $\theta=1^\circ$; top row $\sigma^+$, bottom row $\sigma^-$.  All panels
share one logarithmic color scale spanning five decades below the global
maximum over all states and channels, and intensities are directly
comparable across panels.  The dashed
circle marks $|\qv|=\GM$.  Far-field optics reads only the central point of
each panel---where, within the $C_3$-symmetric leading-harmonic model, the
$m=+1$ state $p_+$ (like $d_-$, not shown) vanishes exactly---while graphene at spacer
$z$ reads the longitudinal combination [Eq.~\eqref{eq:longitudinal}] over the disk
$q\lesssim q_{\mathrm{pk}}=3/(2\gamma z)$.}
\label{fig:pmaps}
\end{figure*}

Figure~\ref{fig:pmaps} shows the full
momentum maps $|P_n^{(\pm)}(\qv)|^2$.  To read them, each panel's center is
$\qv=0$---the single point far-field optics samples---brightness is the log
intensity on one common scale, and the top and bottom rows are the two circular
channels $\sigma^+$ and $\sigma^-$.  A state emits $\sigma^\pm$ photons
only when the center of its map is bright in that row, as the selection
rule of Eq.~\eqref{eq:smearing} states.  Graphene reads the off-center
weight inside the dashed $\GM$ circle.  The $\sigma^-$ maps are nearly isotropic,
whereas the $\sigma^+$ maps display trefoil nodal patterns, moir\'e satellites,
and---for the $p$ states---central intensities that differ by more than two
orders of magnitude between the chiral partners.  All of this structure is
inaccessible to far-field optics but lies inside the graphene window
$q\lesssim1/z$ for $z$ of a few nanometers [Fig.~\ref{fig:model}(d)].

\begin{figure*}[tbp]
\includegraphics[width=\textwidth]{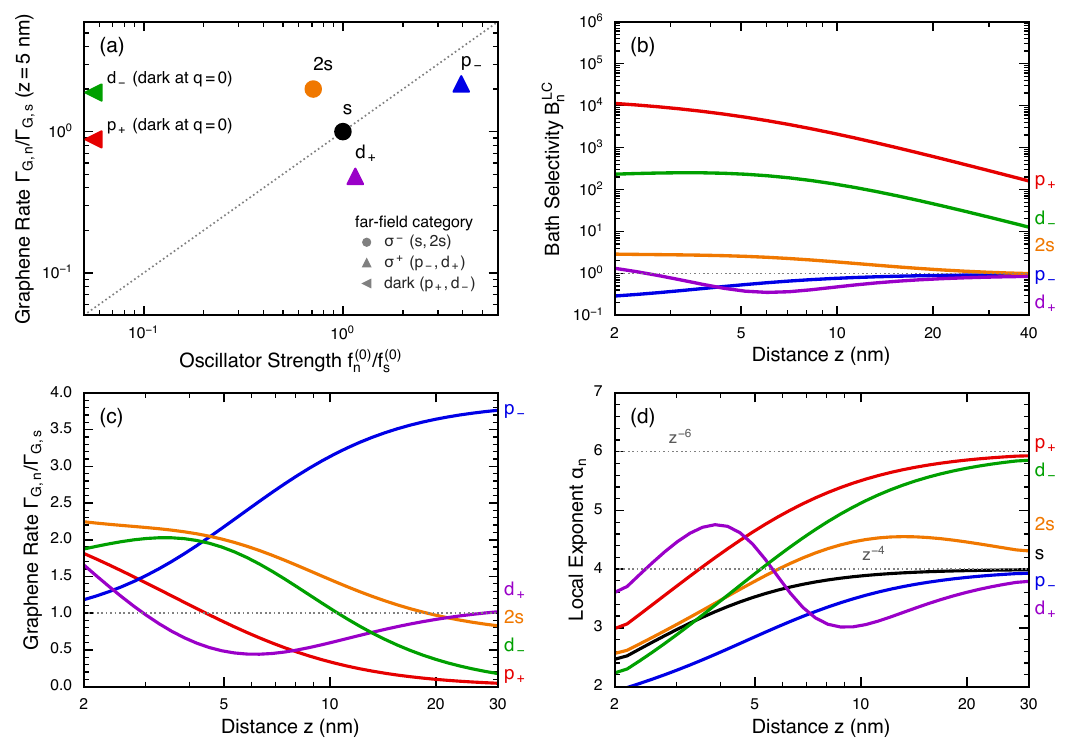}
\caption{State selectivity of the two baths and its distance dependence.
(a)~Photon versus graphene brightness (both normalized to the $s$
state, $z=5$\,nm, longitudinal kernel).  States above the diagonal are
graphene favored.  The $q=0$-dark $p_+$ and $d_-$ are placed at the left
boundary as left-pointing triangles, since a vanishing normal-incidence oscillator
strength cannot be shown on the logarithmic axis.  The marker shape encodes
the far-field category (circle $\sigma^-\!:s,2s$; triangle $\sigma^+\!:p_-,d_+$;
left-pointing triangle, the $q=0$-dark $p_+,d_-$).
The panel thus doubles as the two-axis classification
(far-field category\,$\times$\,graphene) of Sec.~\ref{sec:rules}.  (b)~Physical
(light-cone-normalized) bath selectivity $B^{\mathrm{LC}}_n(z)$ of the five excited
states (dotted line: the $s$ reference, $B^{\mathrm{LC}}_s=1$), normalizing the graphene
rate by the light-cone-integrated radiative rate $\Gamma^{\mathrm{LC}}_{\gamma,n}$.  The
normal-incidence $B^{(0)}_n$ of Eq.~\eqref{eq:Bn} diverges for the dark states.  The selectivity
$B^{\mathrm{LC}}_n$ is finite for every state and preserves the ranking.  For
the bright states $B^{\mathrm{LC}}_n$
tracks the tabulated $B^{(0)}_n$ (Table~\ref{tab:states}), while the $q=0$-dark
$p_+$ and $d_-$ are ${\sim}10^{2}$--$10^{4}$ times graphene favored at
few-nanometer $z$, decreasing rapidly as the kernel window collapses toward
$q=0$.  (c)~Graphene-rate contrast
$\GG{n}(z)/\GG{s}(z)$ (dotted line: $s$).  The state ordering restructures with distance.
(d)~Local distance exponent $\alpha_n(z)=-\dd\ln\GG{n}/\dd\ln z$.}
\label{fig:select}
\end{figure*}

Table~\ref{tab:states} and Fig.~\ref{fig:select}(a) quantify the
resulting state classification.  Within the $C_3$-symmetric model the two $m=+1$
states emit essentially no far-field light, yet their graphene rates are
of the order of the bright-state rates---$\GG{p_+}(5\,\mathrm{nm}) = 0.88\,\GG{s}$
and $\GG{d_-}(5\,\mathrm{nm})=1.89\,\GG{s}$.  Graphene couples to momentum
content that lies overwhelmingly outside the optical light cone.  The dark $p_+$
ratio is also the clearest signature of the texture phase.  For the opposite sign
$\beta=\pi$ the ratio would be $0.21$, whereas far-field emission samples only
$|P^{(\pm)}_n(0)|^2$ and is blind to the sign.  Referred to their light-cone radiative
tails (Sec.~\ref{sec:rules}), the dark states have large but finite
selectivities, $B^{\mathrm{LC}}_{p_+}(5\,\mathrm{nm})\simeq5.5\times10^{3}$
and $B^{\mathrm{LC}}_{d_-}(5\,\mathrm{nm})\simeq2.3\times10^{2}$,
far above those of the bright states on the light-cone-normalized axis of
Fig.~\ref{fig:select}(b).

Among the bright states the two baths also
disagree.  The $2s$ state is graphene favored ($B^{\mathrm{LC}}=2.6$ at $z=5$\,nm),
whereas $p_-$ and $d_+$ are photon favored ($B^{\mathrm{LC}}=0.53$ and
$0.39$; the tabulated normal-incidence $B^{(0)}_n$ of
Table~\ref{tab:states} track these to within ${\sim}10\%$).  The brightest state ($p_-$)
is the \emph{least} graphene-favored, because its large $\qv\simeq0$
strength enters the denominator of Eq.~\eqref{eq:Bn}.

\emph{Degenerate doublet.}
The chiral labels $p_\pm$ diagonalize $C_3$.  Their \emph{photon} decay matrix
is diagonal at normal incidence because $p_+$ is dark ($P_{p_+}(0)=0$), and
their \emph{graphene} decay matrix is diagonal because the partners belong to
different $C_3$ classes (cross element below $10^{-5}$ of the diagonal;
Appendix~\ref{app:numerics}).  Thus, the two partners decay at
distinct rates, with the ratio of the bright to the dark decay rate
$\Gamma_{--}/\Gamma_{++}=2.7$ at $z=5$\,nm and $16$ at
$10$\,nm.  In the ideal $C_3$-symmetric model the partners are degenerate and form one
line.  In a device the protecting splitting separates them into two resolved
lines (Sec.~\ref{sec:protocol}).

\begin{figure*}[tbp]
\includegraphics[width=\textwidth]{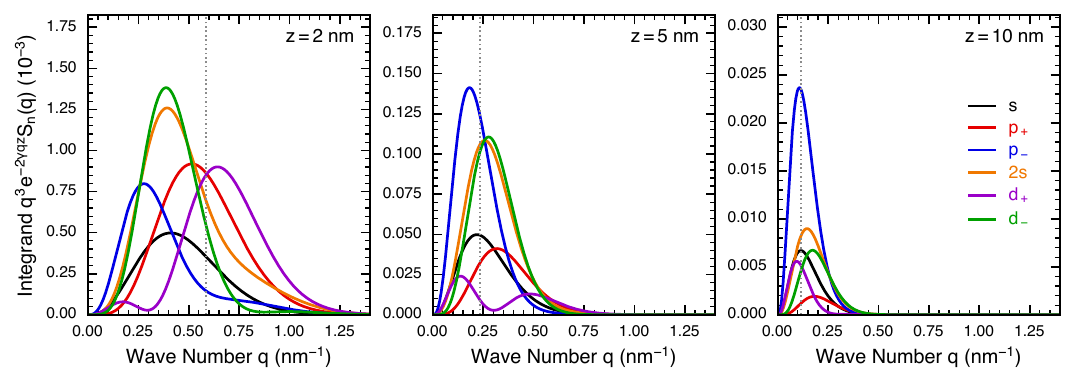}
\caption{Mechanism of the distance-driven ordering reversal.  The graphene-rate
radial integrand $q^3e^{-2\gamma qz}S_n(q)$ for the six moir\'e states (AA, $\theta=1^\circ$) at
$z=2$, $5$, and $10$\,nm (dotted line: the kernel peak $q_{\mathrm{pk}}=3/(2\gamma z)=0.59$,
$0.23$, and $0.12$\,nm$^{-1}$).  The area under each curve is proportional to $\GG{n}(z)$
with a common prefactor.  The per-state $\omega_n$ factors of
Table~\ref{tab:states}, omitted here, shift the areas by only a few percent.  As $z$ grows
the window slides to small $q$.  At $z=2$\,nm the high-$q$-weighted states
($2s$, $d_\pm$, and $p_+$) dominate, whereas by $z=10$\,nm (note
the ${\sim}60\times$ smaller vertical scale) the small-$q$ bright $p_-$ and $2s$
dominate and the $q=0$-dark $p_+$ is strongly suppressed.}
\label{fig:integrand}
\end{figure*}

The distance dependence adds a second, independent discriminator
[Fig.~\ref{fig:select}(c)].  At $z=2$\,nm the graphene kernel reaches deep
into the momentum maps and favors the states with more weight at large $q$
($d_\pm$, $2s$, and $p_+$).  As $z$ grows the kernel window narrows onto small
$q$ and the ordering restructures.  The mechanism is explicit in the radial
integrand $q^3e^{-2\gamma qz}S_n(q)$, whose area over $q$ is $\GG{n}(z)$
[Fig.~\ref{fig:integrand}].  The window
$q_{\mathrm{pk}}=3/(2\gamma z)$ slides from $0.59$ to $0.12$\,nm$^{-1}$ over
$z=2$--$10$\,nm, transferring weight from the high-$q$-heavy dark states to the
small-$q$ bright states.

Concretely, the dark $p_+$ crosses below $s$ at
$z\simeq4.5$\,nm and the bright $d_+$ at $z\simeq3.0$\,nm.  The dark $d_-$
stays above $s$ out to $z\simeq10.5$\,nm.  The bright $p_-$ remains above $s$
throughout (the small-$q$ weight of $p_-$ tracks its large oscillator strength), whereas
the bright $2s$, favored at small $z$ by its high-$q$ weight (negative low-$q$ curvature,
Sec.~\ref{sec:validity-gauss}), falls
below $s$ only at $z\simeq18.9$\,nm on its way to the far-field ratio
$f^{(0)}_{2s}/f^{(0)}_s\simeq0.7$.  The local exponents at $z=10$\,nm range from
$\alpha_{d_+}=3.0$ to $\alpha_{p_+}=5.5$
[Fig.~\ref{fig:select}(d), Table~\ref{tab:states}].  The spread of more than
two units departs from the universal $\alpha=4$ of any point-dipole theory and
from the ideal multipole sequence $\alpha=4+2|m|$ of
Appendix~\ref{app:toy}, which finite size, dipole texture, and $C_3$
mixing renormalize.

Measuring these exponents requires resolving the rate over a decade in $z$.
Because the dark states' rates fall fastest, the spread is most cleanly
resolved at $z\simeq3$--$6$\,nm, and the $z=10$\,nm values are characteristic
reference values.

\section{Twist-angle control}
\label{sec:control}

\begin{figure*}[tbp]
\includegraphics[width=\textwidth]{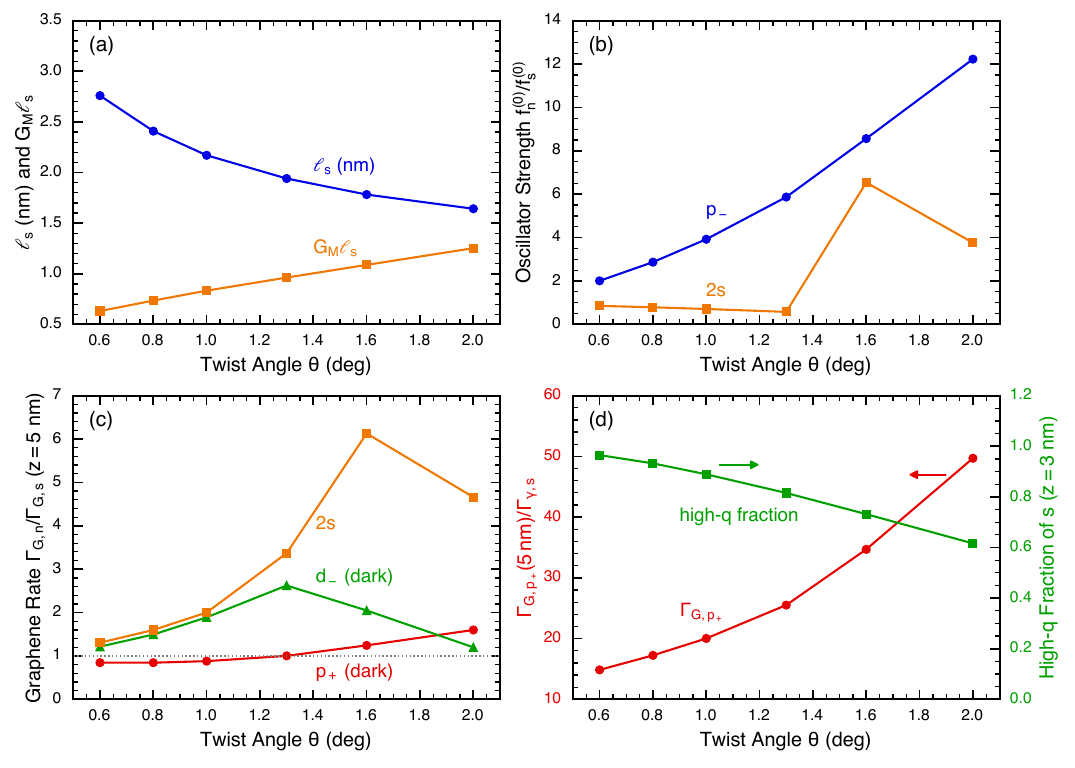}
\caption{Twist control of the state-selective coupling (AA stacking; moir\'e
potential and texture parameters held fixed).  (a)~Orbital size $\ell_s$ and
$\GM\ell_s$.  (b)~Oscillator strengths of the bright $p_-$ and $2s$ relative to $s$.
(c)~Graphene state contrast at $z=5$\,nm (the $p_+$ and $d_-$ curves are
the $q=0$-dark states).  (d)~Absolute graphene rate $\GG{p_+}/\Gph{s}$ of the dark $p_+$ state at $z=5$\,nm (left
axis) and the high-momentum ($q>\GM/2$) fraction of the $s$-state rate at $z=3$\,nm
(right axis).}
\label{fig:control}
\end{figure*}

Because the corner-sampling radius $\GM/\sqrt3$ and the orbital momentum
spread $1/\ell_s$ scale differently with twist
[$\ell_s\propto\aM^{1/2}$ in the harmonic-well regime,
Fig.~\ref{fig:control}(a)], the entire
classification is twist tunable (Fig.~\ref{fig:control}).  Within the
$C_3$-symmetric model the exact selection rules themselves are angle
independent, with $p_+$ and $d_-$ dark at $\qv=0$ at every $\theta$.  However, the
strengths and contrasts vary over wide ranges between $\theta=0.6^\circ$ and $2^\circ$.

The $p_-$
oscillator strength grows from $2.0$ to $12$ times the $s$ strength as
the corner
samples climb the $p$-envelope maximum
[Fig.~\ref{fig:control}(b)], and the graphene contrast of the dark
$p_+$ state rises from $0.85$ to $1.6\,\GG{s}$ [Fig.~\ref{fig:control}(c)].
The dark $d_-$ contrast peaks at $2.6$ near $\theta\simeq1.3^\circ$ and the
$2s$ contrast at $6.1$ near $1.6^\circ$.  The absolute dark-state scale
$\GG{p_+}(5\,\mathrm{nm})/\Gph{s}$ grows from $15$ to $50$, and the
high-momentum weight ($q>\GM/2$) decreases from $0.97$ to $0.62$ as the
moir\'e momentum scale grows past the kernel window
[Fig.~\ref{fig:control}(d)].  Twist angle is thus
a means of selecting \emph{which} hidden state graphene reads most
strongly, within the rigid-lattice parameterization (lattice
reconstruction at small twist is outside the model).

\section{Generalized radii, single-radius limit, and gate control}
\label{sec:validity-gauss}

At low momentum, graphene measures a single quantity.  The quantity entering
Eq.~\eqref{eq:graphene} after angular integration is the longitudinal
spectral density
\begin{equation}
  S_n(q)=\bigl\langle|\hat q\cdot\Pn{n}(\qv)|^2\bigr\rangle_{\varphi_q},
  \qquad
  S_n(q)\simeq S_n(0)\Bigl[1-\frac{\Reff^2q^2}{2}\Bigr],
  \label{eq:Sq}
\end{equation}
whose low-$q$ curvature defines a \emph{generalized}
transition-polarization radius through the \emph{signed} coefficient
$\Reff^{2}$.  For $\Reff^{2}>0$, $\Reff$ is the radius that a distance sweep
determines, while $\Reff^{2}<0$ means that no real Gaussian radius can
reproduce the low-$q$ behavior.  For a phase-uniform Gaussian source
$\Reff$ reduces to the $R_X$ of Ref.~\onlinecite{Wakabayashi2026quench}.
The computed radii expose the difference between graphene's view and real
space.  At $\theta=1^\circ$ the $s$ state has $\Reff=1.8$\,nm, below its
orbital size $\ell_s=2.17$\,nm because the corner components add finite-$q$
weight that partly cancels the envelope curvature (for the opposite sign
$\beta=\pi$ the interchannel interference would even reverse the sign of the
curvature).  The $p_-$ and $d_+$ states have $\Reff=4.4$ and $7.8$\,nm, and the $2s$ state has
$\Reff^{2}<0$.  For the $q=0$-dark $p_+$
and $d_-$ states $S_n(0)=0$ identically---their spectral weight sits entirely at finite
momentum [$p_+$ in Fig.~\ref{fig:gauss}(b)].  Thus, the single-radius description is
undefined for them.

Where $\Reff^2>0$ the large-$z$ expansion
$\GG{}\propto(\gamma z)^{-4}[1-\tfrac52\Reff^2/(\gamma z)^2+\cdots]$ reproduces
the leading correction of Ref.~\onlinecite{Wakabayashi2026quench} with $\Reff$ in
place of $R_X$.  The theory of Ref.~\onlinecite{Wakabayashi2026quench} is then
the controlled large-$z$ limit for bright states with positive low-$q$
curvature, and the radius extracted from a distance sweep is the
graphene-weighted $\Reff$, distinct from the orbital size.  At $\theta=2^\circ$
even the $s$ state has $\Reff^2<0$ (Appendix~\ref{sm:gauss}).

Replacing the neutral interband kernel by the $T=0$ Dirac-RPA loss function of
doped graphene~\cite{Ando2006,Wunsch2006,Hwang2007} preserves the
state-selection conclusions, and gating the Fermi level suppresses the transfer, with the computed
normalized gate curves of $s$ and $p_+$ close to each other, from $T=0$ up
to room temperature.  Because the reachable
modulation is only a few percent ($|\mu_F|\lesssim0.5$\,eV, whereas full
switch-off needs $|\mu_F|\gtrsim0.68$\,eV), gate control serves as an auxiliary
calibration tool (Supplemental
Sec.~S1).

\section{Scope and experimental protocol}
\label{sec:protocol}

\emph{Absolute branching ratio and electromagnetic environment.}
Because both rates derive from one $\Pn{n}$, the branching ratio
$Q_n(z)=\GG{n}(z)/\Gph{n}$ of a bright state is free of the transition-dipole
scale $\mathcal{D}'$ and of the relative-motion contact amplitude.  With the
universal interband loss of graphene and the radiative rate of this in-plane
source embedded in hBN, $Q_n$ takes the closed form of Eq.~\eqref{eq:Q}, which
contains only known constants (Appendix~\ref{app:derivation}).
Equation~\eqref{eq:Q} is a
\emph{dipole-scale-free homogeneous-hBN benchmark}.  In the point-dipole limit it reduces to $Q=(d_0/z)^4$ with
$d_0=11.6$\,nm at $\hbar\omega=1.35$\,eV, the order of the illustrative
$d_0\simeq15$\,nm of Ref.~\onlinecite{Wakabayashi2026quench}.  In homogeneous
hBN the graphene channel exceeds the bare radiative rate out to
$z\simeq11.5$--$13$\,nm, with $Q_s(5\,\mathrm{nm})\simeq23$ and
$\GG{p_+}(5\,\mathrm{nm})\simeq20\,\Gph{s}$ (Supplemental
Sec.~S3).

\emph{Isolating the graphene rate.}
A measured lifetime contains all decay channels,
$\tau_n^{-1}=\Gph{n}+\GG{n}+\Gamma_{\mathrm{nr},n}+\sum_mW_{n\to m}$, with the
nonradiative rate $\Gamma_{\mathrm{nr},n}$ and the interlevel relaxation rates
$W_{n\to m}$, and we predict only the graphene and radiative parts.  The graphene part is isolated by a spacer
series at fixed gate, with the $z$-dependent backgrounds calibrated, or by a
gate-referenced difference $\Delta\Gamma_n(z)=\eta_n\GG{n}(z)$ between low and high
$|\mu_F|$, where $\eta_n$ is the fractional gate modulation of the graphene rate.
Because the reachable gate modulation is only a few percent, the gate difference amplifies
gate-induced background changes and requires stringent control and independent
calibration of the nonradiative, energy, and detection changes at fixed
displacement field~\cite{Wakabayashi2026quench} (Supplemental
Sec.~S6).

\emph{Conditional readout of the dark partner.}
In a device the dark partner survives $C_3$ breaking only under a protecting
splitting $\Delta_B\ge9.1\,\delta_{\rm split}$, where $\delta_{\rm split}$ is the bare
doublet splitting induced by the $C_3$-breaking perturbation
(Appendix~\ref{sm:strain}).  A protecting splitting of the order of $1$\,meV
exceeds the ${\sim}0.1$\,meV linewidth~\cite{Seyler2019}, and $p_-$ and $p_+$
are then two resolved lines, and $\GG{p_+}$ is read from the
lifetime of the weakly bright $p_+$ line, provided $\Delta_B\gtrsim5k_BT$.  The
readout must be polarization and valley resolved, since the field also splits
the time-reversed $-K$ partners, and sensitive, since the line's
photoluminescence yield is small.  The small radiative and fast-partner
corrections follow from the measured oscillator strength of the $p_+$ line (Supplemental
Sec.~S5).

\emph{Protocol.}
The measurement uses one heterobilayer with stepped hBN spacers and a compensated graphene gate.
(i)~Identify the moir\'e lines spectroscopically~\cite{Seyler2019,Tran2019,Jin2019,Choi2021}
and tag their polarization.  (ii)~For each line $n$ and step $z$, measure
$\Delta\Gamma_n(z)$, applying the protecting field and resolving the doublet into
two lines to address $p_+$.  (iii)~Form the ratios of Table~\ref{tab:states}.
Table~\ref{tab:measurement} maps each experimental control to its raw
observable, the derived quantity, and the prediction it tests.  The predicted signatures are the ratios
$\GG{n}/\GG{s}$ of Table~\ref{tab:states}, the crossings of the reconstructed
$\GG{n}(z)=\Delta\Gamma_n(z)/\eta_n$ curves ($\eta_n$ is state dependent) at $z\simeq3.0$ ($d_+$), $4.5$ ($p_+$), $10.5$ ($d_-$),
and $18.9$\,nm ($2s$) in homogeneous hBN, the spread of local slopes over
$\alpha\simeq3.0$--$5.5$ at $z=10$\,nm, and the twist evolution of
Fig.~\ref{fig:control}.  With protection, the resolved $p_+$ line adds a
graphene-induced rate of order $20\,\Gph{s}$ at $z=5$\,nm
although it is nearly dark in emission.

\begin{table*}[tbp]
\caption{\label{tab:measurement}%
Observable roadmap: which experimental control isolates which prediction.
Grade \emph{model-intrinsic} denotes a prediction independent of the
transition-dipole prefactor.  Grade \emph{device-dependent} denotes a prediction that requires an
independently calibrated electromagnetic environment or gate response.
Grade \emph{conditional} denotes a prediction that holds only under a protecting $C_3$-preserving splitting.
Derived quantities are inferred from raw observables under stated assumptions;
$\Gamma_{\mathrm{nr}}$ and the interlevel relaxation rates are calibrated
inputs.
$^{\rm a}$Identical $z$, dielectric environment, temperature, and excitation across
devices.  $^{\rm b}$Reachable modulation
$\eta_n\simeq0.03$--$0.07$, which requires stringent control of gate-induced
backgrounds (Supplemental Sec.~S6).  A spacer series with calibrated
$z$-dependent backgrounds is an alternative.  $^{\rm c}$Homogeneous-hBN benchmark $Q_s(5\,\mathrm{nm})\simeq23$
(Supplemental Sec.~S3); device stacks shift $Q_n$ more
strongly than the state ratios (Appendix~\ref{app:robust}).  $^{\rm d}$Requires
$\Delta_B\ge9.1\,\delta_{\rm split}$ and $\Delta_B\gtrsim5k_BT$ (Appendix~\ref{sm:strain};
Supplemental Sec.~S5).}
\newcommand{\rl}{\raggedright\hspace{0pt}}
\begin{ruledtabular}
\begin{tabular}{p{0.105\textwidth}p{0.175\textwidth}p{0.215\textwidth}p{0.215\textwidth}p{0.20\textwidth}}
\rl Grade & \rl Experimental control & \rl Raw observable & \rl Derived quantity &
\rl Prediction tested\tabularnewline
\hline
\rl model-intrinsic & \rl compare lines $n$ at fixed $z$ & \rl line-resolved decay
rates under a common isolation protocol$^{\rm b}$ (gate or spacer series);
polarization &
\rl $\GG{n}/\GG{s}$ (Table~\ref{tab:states}) & \rl state ranking by finite-$q$
spectral weight\tabularnewline
\rl model-intrinsic & \rl stepped hBN spacer $z$ & \rl $\tau_n^{-1}(z,\mu_F)$$^{\rm b}$ &
\rl $\GG{n}(z)$, crossings, $\alpha_n(z)$; $\Reff$ (bright, large $z$) &
\rl a broad momentum window that slides with $z$\tabularnewline
\rl model-intrinsic & \rl twist $\theta$ (separate devices)$^{\rm a}$ & \rl the above,
in each device & \rl $\theta$ dependence of ratios and crossings
(Fig.~\ref{fig:control}) & \rl corner radius $\GM/\sqrt3$ versus orbital
size\tabularnewline
\rl device-dependent & \rl graphene gate $\mu_F$ & \rl $\tau_n^{-1}$ at ${\ge}2$
gate settings & \rl $\Delta\Gamma_n=\eta_n\GG{n}$$^{\rm b}$ & \rl isolation of
$\GG{n}$ from the background\tabularnewline
\rl device-dependent & \rl radiative calibration & \rl polarization-resolved
absorption or reflectance, photoluminescence excitation, or
population-calibrated photoluminescence & \rl $f^{(0)}_n$;
$\Gph{n}$ (with a photon-environment model); $Q_n$$^{\rm c}$ & \rl absolute
branching ratio\tabularnewline
\rl conditional & \rl identify or tune a $C_3$-preserving splitting $\Delta_B$ &
\rl energies, polarization, and lifetimes of the resolved lines versus field &
\rl $\Delta_B$, $\delta_{\rm split}$; corrected $\GG{p_+}$$^{\rm d}$ &
\rl graphene coupling of the $q=0$-dark partner\tabularnewline
\end{tabular}
\end{ruledtabular}
\end{table*}

\section{Conclusion}
\label{sec:conclusion}

We computed the microscopic transition polarization $\Pn{n}(\qv)$ of moir\'e
interlayer excitons from the moir\'e continuum model and showed that far-field
emission and graphene transfer, evaluated from this single source through its
transverse ($q\simeq0$) and longitudinal (finite-$q$) projections, select
different exciton states.  The graphene-rate ratios differ from the optical ones
and their ordering inverts with spacer thickness.  Within the $C_3$-symmetric
leading-harmonic model, two of the six lowest states---the chiral $p_+$ and
$d_-$---are in addition dark at normal incidence yet transfer to graphene at
rates comparable to that of the bright ground state.  In a device this
additional prediction requires a splitting that protects the chiral partners
(Appendix~\ref{sm:strain}).  Their graphene rates depend
on the sign of the texture phase, a quantity invisible in far-field emission,
which a first-principles calculation determines ($\beta=0$).  Thus, the
dark-partner contrast also tests this sign.  The $q=0$ selection rules
recover the classification of Ref.~\onlinecite{Wu2018}.  The new element is the
finite-$q$ structure obtained
by constructing each $\Pn{n}(\qv)$ from the registry-dependent vector dipole
texture and using each $\Pn{n}$ for both baths---the state-,
polarization-, and registry-resolved graphene response and its distance-driven
ranking inversion.

The prefactor-independent predictions are the state ratios, their ordering
inversion, the local distance exponents, the graphene-weighted radius $\Reff$,
and their twist dependence.  The absolute branching scale and
specific crossing distances depend on the device environment
(Appendix~\ref{app:robust}).

Experimentally, the graphene contribution can be isolated through a calibrated
spacer series or a gate-referenced lifetime difference, and the dark partner
appears as a separate, weakly bright line once the protecting splitting is
resolved.

Graphene proximity is thus a momentum-selective near-field filter of
moir\'e-exciton structure: a coarse-grained, Laplace-type momentum
spectroscopy motivated by recent momentum-resolved transfer
experiments~\cite{Moghe2026}.

\begin{acknowledgments}
This work was supported by JSPS KAKENHI (Grants No.~JP25K01609 and
No.~JP22H05473).  The first-principles calculations were carried out in part on the
Numerical Materials Simulator at the National Institute for Materials Science.
\end{acknowledgments}

\section*{Data availability}
The data used and analyzed during the current study are available from the
corresponding author upon reasonable request.

\appendix

\section{From the WLM matrix element to the finite-\texorpdfstring{$q$}{q} transition
polarization}
\label{app:derivation}

WLM derive, in a two-center local approximation valid for long moir\'e
periods, the interlayer optical \emph{current} matrix element
$\bm J(\bm r_0)$ as a function of the local interlayer displacement
$\bm r_0$, with the three-harmonic form quoted in Eq.~\eqref{eq:texture}
[Eqs.~(9)--(13) of Ref.~\onlinecite{Wu2018}].  In a twisted bilayer the
displacement varies slowly, $\bm r_0\to\theta\,\hat z\times\Rv$, which maps
the monolayer reciprocal vectors $\bm G^{(n)}$ of the WLM expansion onto the
moir\'e vectors $\bm b_n=\theta\,\bm G^{(n)}\times\hat z$.  The local matrix
element becomes the moir\'e-\emph{covariant} optical matrix-element field
$\bm d(\Rv)$ of
Eq.~\eqref{eq:texture}---only the undressed
$e^{-\ii\bm c\cdot\Rv}\bm d(\Rv)$ is strictly periodic, while $\bm d$ itself acquires a Bloch
phase under moir\'e translations---with the Bloch-scale physics (atomic orbitals,
interlayer tunneling, relative-motion contact amplitude) absorbed into the
constants $D_\pm$.  Explicitly, the dipole amplitude is defined from the
WLM current amplitude by $\bm d(\Rv;\omega_n)=\bm J(\Rv)/(-\ii\omega_n)$,
evaluated at the transition frequency $\omega_n$ of the state considered.

Three scales separate.  They are the lattice constant (absorbed into $D_\pm$), the
electron--hole relative motion ($a_X\simeq1.3$\,nm, restored as
$F_{\mathrm{rel}}$ in Appendix~\ref{app:robust}), and the COM/moir\'e scale
($\ell_s$, $\aM$) treated explicitly.  On the last scale the interband
coherence of exciton state $n$ has the transition-current density
$\bm j_X(\Rv)=-\ii\omega\,\bm d(\Rv)W_n(\Rv)\equiv-\ii\omega\Pn{n}(\Rv)$,
which identifies $\Pn{n}(\Rv)=\bm d(\Rv)W_n(\Rv)$ as the transition
\emph{polarization} density.  Its Fourier transform is Eq.~\eqref{eq:Pq}.
The continuity equation for the oscillating coherence,
$\nabla\cdot\bm j_X=-\partial_t\rho_X$, then gives the transition charge
\begin{equation}
  \rho_X(\qv)=-\ii\,\qv\cdot\Pn{n}(\qv),
  \label{eq:continuity}
\end{equation}
i.e., the transition charge comes only from the longitudinal projection of $\Pn{n}$.  The
golden-rule rate for Coulomb coupling of this charge to graphene's density
fluctuations at distance $z$,
$\Gamma\propto\int\dd^2q\,v_q^2e^{-2\gamma qz}|\rho_X(\qv)|^2
[-\mathrm{Im}\,\chi_{\mathrm{G}}(q,\omega)]$ with
$v_q=2\pi e^2/(\varepsilon_{\mathrm{env}} q)$
[$\varepsilon_{\mathrm{env}}=\sqrt{\varepsilon_\parallel\varepsilon_\perp}\simeq3.9$ the
screening permittivity of hBN], reduces with the small-$q$ interband loss
$-\mathrm{Im}\,\chi\propto q^2/\omega$ to Eq.~\eqref{eq:graphene}.  The full
RPA loss replaces the last factor in Sec.~\ref{sec:validity-gauss}.

Finally, the momentum bookkeeping.  In the frame where the COM kinetic
energy is $\hbar^2q^2/2M_X$ with minimum at $q=0$ [WLM Eq.~(18)], the
bright momenta are the three MBZ corners $\bm q_1,\bm q_2,\bm q_3$ [WLM
Eqs.~(3) and (8)], which have equal kinetic energy and whose centroid is
the kinetic minimum.  The in-plane momentum $\qv$ transferred to the
photon or to graphene is the \emph{true} transition-density momentum.  Thus,
the texture harmonics must sit on this $C_3$-closed corner star---hence
the dressing $e^{+\ii\bm c\cdot\Rv}$ of Eq.~\eqref{eq:texture}, which places
the harmonics at $\{-\bm q_1,-\bm q_2,-\bm q_3\}$.

WLM's
footnote~40 defines the alternative frame in which the optically active
momenta are moved to zero (texture harmonics at $\{0,\bm b_a,\bm b_b\}$).
In that frame the kinetic term is $\hbar^2|\qv+\bm q_1|^2/2M_X$, and the
two frames are related by the unitary plane-wave shift.  Mixing the
frames---the undressed texture with the unshifted kinetic term---breaks
the $C_3$ covariance of the coupling at $O(1)$ and generates spurious
selection-rule smearing and a nonzero degenerate-doublet decay-matrix cross
element.  With the consistent choice used here the smearing vanishes and the
cross element is zero by $C_3$ (Appendix~\ref{app:numerics}).

\emph{Absolute scale of the branching ratio.}
Fermi's golden rule for the transition charge density $\rho_X$ (in units of the
elementary charge $q_e$) coupled to the graphene density by the screened Coulomb
interaction $v_q=2\pi e^2/(\varepsilon_{\mathrm{env}}q)$
($e^2\equiv q_e^2/4\pi\varepsilon_0$) gives
$\GG{n}=(2\pi/\hbar)\int\!\tfrac{\dd^2q}{(2\pi)^2}\,v_q^2 e^{-2\gamma qz}
|\rho_X(\qv)|^2\,S_{\mathrm G}(q,\omega_n)$, with $S_{\mathrm G}$ the graphene
dynamic structure factor per unit energy and area.  The $T=0$
fluctuation--dissipation relation $S_{\mathrm G}=-(1/\pi)\,\mathrm{Im}\,\chi_{\mathrm G}$
ties it to the retarded density response $\chi_{\mathrm G}$.  With the universal
interband loss $-\mathrm{Im}\,\chi_{\mathrm G}=q^2/(4\hbar\omega_n)$
(equivalently $\sigma_1=e^2/4\hbar$) and $|\rho_X|^2=q^2|\hat q\cdot\Pn{n}|^2$
[Eq.~\eqref{eq:continuity}] this yields
\begin{equation}
  \GG{n}(z)=\mathcal C_{\mathrm G}\!\int_0^{q_c}\!\!\dd q\,q^3\,
  e^{-2\gamma q z}\,S_n(q),\qquad
  \mathcal C_{\mathrm G}=\frac{\pi e^4}{\varepsilon_{\mathrm{env}}^2\hbar^2\omega_n},
  \label{eq:GGabs}
\end{equation}
with $S_n(q)=\langle|\hat q\cdot\Pn{n}|^2\rangle_{\varphi_q}$
[Eq.~\eqref{eq:Sq}] and $\mathcal C_{\mathrm G}=C_{\mathrm G}/2\pi$ in terms of the
prefactor of Eq.~\eqref{eq:graphene}.  Dividing by the radiative rate
$\Gph{n}=n_r\,\omega_n^3|\bm\mu_n|^2/(3\pi\varepsilon_0\hbar c^3)$ of this
source, with $\bm\mu_n=q_e\Pn{n}(0)$ and
$n_r=(3\varepsilon_\parallel+\varepsilon_\perp)/(4\sqrt{\varepsilon_\parallel})\simeq2.01$, gives
\begin{align}
  Q_n(z)&=\frac{3\pi\alpha_{\mathrm{fs}}}{4\,\varepsilon_{\mathrm{env}}^2 n_r}
  \Bigl(\frac{c}{\omega_n}\Bigr)^{\!4}
  \frac{\int_0^{q_c}\dd q\,q^3 S_n(q)e^{-2\gamma qz}}
       {|P^{(+)}_n(0)|^2+|P^{(-)}_n(0)|^2},\label{eq:Q}\\
  d_0&=\Bigl(\tfrac{9\pi\alpha_{\mathrm{fs}}}{64}\Bigr)^{1/4}
  \frac{c/\omega}{(\varepsilon_{\mathrm{env}}^2 n_r)^{1/4}\gamma},
  \label{eq:QandD0}
\end{align}
in which the transition-dipole scale and the relative-motion contact amplitude
common to $\GG{n}$ and $\Gph{n}$ cancel identically
($\alpha_{\mathrm{fs}}=e^2/\hbar c$ is the fine-structure constant).

The second line is the structureless
point-dipole limit ($S_n$ constant, $q_c\to\infty$), $Q=(d_0/z)^4$, which has the
form of the in-plane dipole quenching law of Ref.~\onlinecite{Wakabayashi2026quench}.
At $\hbar\omega=1.35$\,eV it gives $d_0=11.6$\,nm, in the $10$--$20$\,nm range of
quenching distances reported for near-field transfer from proximal emitters to
graphene~\cite{Gaudreau2013,Tielrooij2015,Froehlicher2018}.  The
\emph{overall scale} has an $O(1)$ (factor-${\sim}2$) uncertainty of
purely \emph{physical-modeling} origin: the homogeneous-hBN benchmark, the
single source plane, and the small-$q$ form of the loss function
(Table~S1).  Determining the scale to
better than a factor ${\sim}2$
requires the full layered electromagnetic Green tensor.  Thus, we
read $Q_s(5\,\mathrm{nm})\sim23$, $d_0\sim12$\,nm, and the unity
crossings as \emph{order-of-magnitude} benchmarks.  The parametric scalings
($Q\propto\omega^{-4},\,\varepsilon_{\mathrm{env}}^{-2},\,n_r^{-1}$), the
state ratios, the distance exponents, and the orderings are independent of the
overall coefficient.

\section{Localized states: gauge, single-well benchmark, and localization
mechanisms}
\label{app:localization}

\emph{Construction and gauge.}
On a $\Gamma$-centered $18\times18$ $k$ mesh we fix the Bloch gauge by
projection onto harmonic trial orbitals centered at the potential minimum
(Gaussian for $m=0$ bands; chiral $p_\pm$/$d_\pm$ pairs with L\"owdin
orthonormalization for the degenerate doublets).  A Wannier packet is
gauge dependent in principle; two facts make the present ones physically
meaningful.  First, varying the trial width by $\pm30\%$ changes the
cancellation-sensitive contrasts and exponent of Appendix~\ref{app:numerics},
item~(vii), by less than $10^{-5}$.  For these isolated flat bands
the projection phases are insensitive to the trial.

Second,
a \emph{gauge-free} construction---solving one moir\'e trap directly by
finite differences in an open box and forming the chiral combinations of the
two degenerate $p$ solutions---reproduces the full observable set.  The
$s$--$p$ spacing ($19.2$\,meV) matches the miniband value, the exact
photon zeros are reproduced at the $10^{-5}$ level determined by the box, and
the dark-state contrast $\GG{p_+}/\GG{s}=0.816$--$0.873$ at $z=5$\,nm versus
$0.881$ ($\beta=0$), with all order-unity
photon and graphene ratios agreeing to within $8\%$ across box sizes of
$0.40$--$0.50\,\aM$.  The construction-sensitive entries are the exact
zeros themselves ($10^{-5}$ of the $s$ strength in the box, versus
$10^{-8}$ in the Bloch route).  Each construction resolves them down to its
own numerical floor.  The flat-band orbitals \emph{are} the
single-trap bound states.

\emph{Localization mechanisms and coherence.}
An ideal periodic system has Bloch eigenstates; a localized packet must be
prepared and maintained.  The computed bandwidths $w$ at $\theta=1^\circ$ are
$0.05$, $2$, and $49\,\mu$eV for the $s$, $p$, and $2s$ manifolds,
corresponding to intersite tunneling times $\hbar/w\simeq12$\,ns, $0.3$\,ns,
and $13$\,ps.  Static disorder and strain far exceed the two lowest
bandwidths in current samples---moir\'e excitons appear experimentally as
quantum-emitter-like localized lines~\cite{Seyler2019,Baek2020}---and for
$z\lesssim4$\,nm the graphene channel itself
[$\GG{}^{-1}=\tau_{\gamma}/Q_n\sim10$\,ps with $\tau_\gamma=\Gph{}^{-1}$ at $z=3$\,nm for a nanosecond
radiative lifetime, Eq.~\eqref{eq:Q}] is comparable to or faster than even
the $2s$ tunneling time.  Thus, the decay
is site local.

In the opposite, fully site-incoherent limit described by
the density matrix $\hat\varrho_X=\sum_{i}p_{i}
|W_{nR_i}\rangle\langle W_{nR_i}|$ (sites $R_i$), the rates are additive over sites and
each site reproduces the single-orbital rates used here, because the
overlap of neighboring-site transition densities is negligible for these
flat bands.  The present results apply to the experimentally
relevant site-localized, quantum-emitter-like lines and to site-incoherent
mixtures of localized states.  Coherently prepared Bloch excitons and
collective intersite radiative or bath-mediated effects are outside the
scope of this work.

\section{Numerical method and validation}
\label{app:numerics}

\emph{Diagonalization and momentum representation.}
Equation~\eqref{eq:H} is solved in the plane-wave basis
$H_{\Gv\Gv'}(\bm k)=\frac{\hbar^2|\bm k+\Gv|^2}{2M_X}\delta_{\Gv\Gv'}
+V_{\Gv-\Gv'}$ with $V_{\pm\Gv_j}=V_0e^{\pm\ii\phi_V}$, on six reciprocal
shells (127 vectors).  The momentum representation follows from the exact
identity $\widetilde W_n(\bm k+\Gv)=\sqrt{A_c}\,\widetilde C_{n\bm k}(\Gv)$
($A_c$ the moir\'e cell area).  The set $\{\bm k+\Gv\}$ is a regular grid of
spacing $\GM/18$ in reduced coordinates, on which $\widetilde W_n$ is
assembled and interpolated bicubically for arbitrary $\qv$.  Rate integrals
use composite Gauss--Legendre quadrature in $q$ with breakpoints at the
kernel peak $3/(2\gamma z)$ and at the moir\'e rings, and a 96-point
periodic
trapezoid in angle, evaluating the longitudinal combination
Eq.~\eqref{eq:longitudinal} on each ring.

\emph{Validation.}
(i)~For Gaussian orbitals the analytic transforms are reproduced to
$10^{-15}$ by the generic fast-Fourier-transform (FFT) route, and the fitted distance exponents of
the $s$, $p$, $d$ toy states are $4.00$, $6.00$, $8.00$ within $1\%$
(Appendix~\ref{app:toy}).
(ii)~With a uniform dipole the $s$-state rate equals the kernel
$2\ell^2I(z,q_c,\sqrt2\ell)$ of Ref.~\onlinecite{Wakabayashi2026quench} to
$4\times10^{-9}$ relative accuracy.
(iii)~In the deep-well limit ($V_0=60$\,meV) the $s$ orbital overlaps the
harmonic Gaussian by $0.99966$, and the $s$--$p$ spacing agrees with the
single-well solver to $0.14$\,meV.
(iv)~The large-$z$ expansion reproduces the analytic coefficient $\tfrac52$
of the $\Reff^2/(\gamma z)^2$ correction to $2\%$
(verified at $\gamma=1$).
(v)~The exact convolution route and the real-space FFT route for
$\Pn{n}(\qv)$ agree to better than $10^{-13}$.
(vi)~In the well-separated regime ($\GM\ell\gg1$) the moir\'e-ring
contribution decays with the saddle-point rate $2q^*(z)$,
$q^*=\GM-\gamma z/\ell^2+5/(4q^*\ell^2)$ (verified at $\gamma=1$),
matching the numerics to $1.4\%$ and
reducing to the pure near-field factor $e^{-2\gamma\GM z}$ for narrow
rings.

(vii)~\emph{Cancellation-sensitive observables.}  The dark-state contrast $\GG{p_+}(5\,\mathrm{nm})/\GG{s}(5\,\mathrm{nm})$,
the contrast
$\GG{2s}/\GG{s}$, and the exponent $\alpha_{p_+}(10\,\mathrm{nm})$ change by
at most $10^{-4}$ (relative) under any single refinement or all refinements
combined, and the exact photon zeros stay below $10^{-8}$ of the $s$ strength.  The refinements are reciprocal shells
$6\to8$, $k$ mesh $18\to30$, angular mesh $96\to192$, and trial width
$\pm30\%$.

(viii)~\emph{Convention invariance.}  Every rate is unchanged under a
basis-free Cartesian evaluation of $|\hat q\cdot\Pn{}|^2$ in place of the
$\bm e_\pm$ kernel ($4\times10^{-16}$), under global frame rotations
($\leq9\times10^{-8}$), and under real-space origin shifts pushed through
the texture phases (exact).  A full $2\pi$ scan ($24$ points) of the relative
phase of $D_-$ and $D_+$ changes $\GG{s}(5\,\mathrm{nm})$ by up to $18\%$ and
$\GG{p_+}(5\,\mathrm{nm})$ by up to $71\%$.  Symmetry restricts the phase to
$\beta=0,\pi$ (Sec.~\ref{sec:texture}), and
the calculation of Appendix~\ref{sm:dft} gives $\beta=0$.

(ix)~\emph{Corner-star frame checks.}  The dressed vector texture of
Eq.~\eqref{eq:texture} is $C_3$ covariant about the high-symmetry sites to
$10^{-13}$ (about $R_0$, $R_1$, and $R_2$, for AA and AB and for both $\beta$,
to $6\times10^{-14}$) and agrees with the three-harmonic form of
Ref.~\onlinecite{Wu2018} to $3\times10^{-14}$.  The exact photon zeros of
Table~\ref{tab:states} are reproduced below $10^{-8}$ of the $s$ strength.  The
ring-wise harmonic decomposition of $P^{(+)}_nP^{(-)*}_n$ is dominated by the
$l=-2$ harmonic of the class $l\equiv+1$ (mod 3), the $l\equiv-1$ harmonics
staying below $5\times10^{-6}$ of it.  Thus, the interference term of
Eq.~\eqref{eq:longitudinal} survives the angular average, as stated in
Sec.~\ref{sec:kernels}.

(x)~\emph{Degenerate-doublet decay matrices.}  The normal-incidence ($q=0$)
\emph{photon} matrix is diagonal---below $10^{-5}$ of $\Gamma_{\gamma,s}$, its
analytic origin being $P_{p_+}(0)=0$ (the small finite-light-cone radiative
tail of $p_+$ is item~(xi)).  The \emph{graphene} matrix is diagonal as well.
The cross element $\Gamma_{+-}$ is forbidden by $C_3$
(Sec.~\ref{sec:kernels}), and we include it in the Hermitian
$2\times2$ matrix only as a check.  Its \emph{gauge-invariant} magnitude
$|\Gamma_{+-}|/\sqrt{\Gamma_{++}\Gamma_{--}}$, evaluated with the full
longitudinal cross kernel---keeping the interchannel $e^{\pm2\ii\varphi_q}$
terms, i.e.\ \emph{without} assuming channel diagonality---is
$2.0\times10^{-6}$ at $z=2$\,nm, $2.1\times10^{-6}$ at $z=5$\,nm, and
$3.6\times10^{-6}$ at $z=10$\,nm for the interpolated moir\'e Wannier states,
$3\times10^{-14}$ for the $C_3$-projected states on $N_k\times N_k$ meshes with $N_k=18$, $24$, and
$30$, and
$3.5\times10^{-15}$ for analytic harmonic-oscillator chiral states (exact $C_3$,
no interpolation).  Angular averages use the periodic quadrature ($2\pi$ times
the $l=0$ discrete Fourier component), because a non-periodic trapezoid rule on
the periodic $\varphi$ grid would drop the wrap-around interval.

(xi)~\emph{Kernel and light-cone corrections.}  The exact finite-$q$
neutral Dirac response raises $Q_n$ by at most $4\%$ and its RPA
self-screening lowers it by at most $2\%$ (partially canceling; $Q=1$
crossings move by less than $0.01$\,nm).  The bright photon channels vary
by less than $0.4\%$ across the light cone.  The radiative sector uses the
uniaxial ordinary/extraordinary weights $W_o=1/(2k_{z,o})$ on
$q<\sqrt{\varepsilon_\parallel}\,\omega_n/c$ and
$W_e=k_{z,e}/(2\varepsilon_\parallel\omega_n^2/c^2)$ on
$q<\sqrt{\varepsilon_\perp}\,\omega_n/c$ ($k_{z,o/e}$ the out-of-plane
wave numbers of the two branches), reproducing
$(3\varepsilon_\parallel+\varepsilon_\perp)/(4\sqrt{\varepsilon_\parallel})\simeq2.01$
for a flat source (ordinary fraction $0.83$; $9.8\%$ below the isotropic
effective-index value $\sqrt{\varepsilon_\parallel}\simeq2.23$; $n_r$ is an effective
radiative index).  Relative to an isotropic cone the
uniaxial treatment changes the bright-state ratios by less than $0.03\%$ and the
$d_-$ tail by $2\%$, but lowers the $p_+$ tail by a factor ${\sim}2$, because that
tail sits near the cone edge where the extraordinary cutoff acts.  Thus, we quote the $p_+$ tail
to within this factor.  A source-plane shift
$z\to z\pm0.3$\,nm relabels the distance axis, translating all crossings
by $\mp0.3$\,nm.

(xii)~\emph{COM orbital moment.}  Solving one moir\'e trap on an open box (the
gauge-free single-well construction of Appendix~\ref{app:localization}) and forming the chiral
combinations $(\psi_1\pm\ii\psi_2)/\sqrt2$, the $p$ doublet has
$\langle L_z\rangle=\pm0.99$ [with
$\langle L_z\rangle=\langle\psi|-\ii\,\partial_\varphi|\psi\rangle$ evaluated by
finite differences about the trap center], consistent with its $C_3$
eigenvalue $e^{\mp2\pi\ii/3}$ ($m=\pm1$).  Its relation to the doublet splitting
$\Delta_B$ is discussed in Sec.~\ref{sec:com}.

\section{Harmonic-oscillator limits}
\label{app:toy}

For a two-dimensional harmonic well of length $\ell$ the lowest states of
angular momentum $m$ have unit-normalized envelopes
$\psi_m\propto R^{|m|}e^{-R^2/2\ell^2}e^{\ii m\varphi}$ and analytic
transforms
\begin{equation}
  \widetilde\psi_m(\qv)=\frac{2\sqrt{\pi}\,\ell\,(-\ii)^{|m|}}{\sqrt{|m|!}}
  \,(q\ell)^{|m|}e^{-q^2\ell^2/2}e^{\ii m\varphi_q}.
  \label{eq:HOtilde}
\end{equation}
With a uniform dipole, Eq.~\eqref{eq:graphene} gives
$\GG{m}(z)\propto\int q^{3+2|m|}e^{-2\gamma qz}\dd q\propto (\gamma z)^{-(4+2|m|)}$: the
multipole ladder $(\gamma z)^{-4}$, $(\gamma z)^{-6}$, $(\gamma z)^{-8}$ for $s$, $p$, $d$.  These are
the ideal exponents with which we compare the moir\'e results of
Fig.~\ref{fig:select}(d).  Finite size flattens them at
$\gamma z\lesssim3\ell$, and the dipole texture adds a $q$-independent
component
that restores $\alpha\to4$ at large $z$ for every textured-bright state.
The corner-sample amplitudes behind Eq.~\eqref{eq:smearing} follow from
Eq.~\eqref{eq:HOtilde} evaluated at the corner momenta, and the exact
darkness of the $m=+1$ states follows from the pure local angular-momentum
classes of Eq.~\eqref{eq:texture} about the trap.  On the moir\'e lattice
$C_3$ conserves these classes modulo three.  Thus, the selection rules are
exact for the lattice as well.

\section{Generalized radius and the single-radius limit}
\label{sm:gauss}

\begin{figure*}[tbp]
\includegraphics[width=\textwidth]{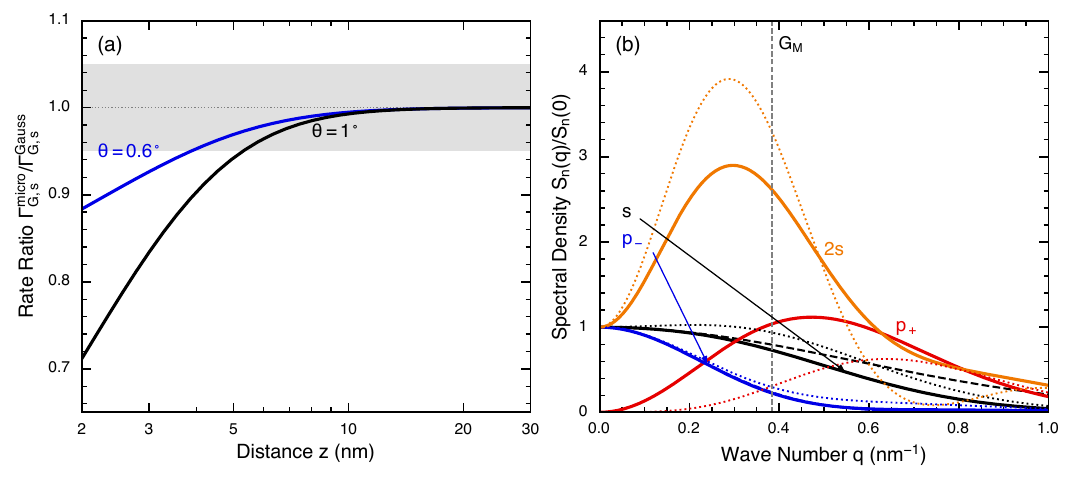}
\caption{(a)~Validity of the single-radius description: microscopic over
Gaussian rate for the $s$ state, with the Gaussian matched to the microscopic
$S(0)$ and $\Reff$ [Eq.~\eqref{eq:Sq}], at $\theta=0.6^\circ$ and $1^\circ$ ($\beta=0$; the
$\beta=\pi$ values are quoted in the text).  The
shaded band marks $\pm5\%$.  At $\theta=2^\circ$, and at $1^\circ$ for $\beta=\pi$,
the $s$ state itself has $\Reff^2<0$, no matched Gaussian exists, and no curve is
drawn.  (b)~Spectral densities $S_n(q)/S_n(0)$ at $\theta=1^\circ$ (solid; dotted:
$\beta=\pi$) and the matched Gaussians of the states with $\Reff^2>0$ (dashed; for $p_-$
the Gaussian is indistinguishable from the microscopic curve).  The
vertical dashed line marks $\GM$, and the $q=0$-dark $p_+$, with $S(0)=0$, is
normalized to the $s$-state $S_s(0)$.}
\label{fig:gauss}
\end{figure*}

The generalized radius $\Reff$ is defined by the low-$q$ curvature of the
longitudinal spectral density, Eq.~\eqref{eq:Sq}.
Figure~\ref{fig:gauss}(a) locates the boundary of the single-radius
theory.  With a coherent Gaussian matched to the microscopic $S(0)$ and $\Reff$, the
microscopic $s$-state rate deviates by a factor $0.71$ at $z=2$\,nm at
$\theta=1^\circ$ ($0.88$ at $0.6^\circ$), converging into the $\pm5\%$
band for $z\gtrsim4$\,nm at $0.6^\circ$ and $z\gtrsim5.3$\,nm at
$1^\circ$ ($\beta=0$).  For $\beta=\pi$ the $s$-state curvature is already
negative at $1^\circ$ (at $0.6^\circ$ the matched Gaussian deviates by $0.63$ at
$2$\,nm and enters the band beyond $6.8$\,nm), and at $\theta=2^\circ$ no real
radius exists for either sign.  Where $\Reff^2>0$, the coefficient
$\tfrac52$ of the large-$z$ expansion quoted in Sec.~\ref{sec:validity-gauss} is
recovered to $2\%$ in the direct numerical check of Appendix~\ref{app:numerics}.

\section{\texorpdfstring{$C_3$}{C3} breaking and the doublet protection window}
\label{sm:strain}

\begin{figure}[htbp]
\includegraphics[width=\columnwidth]{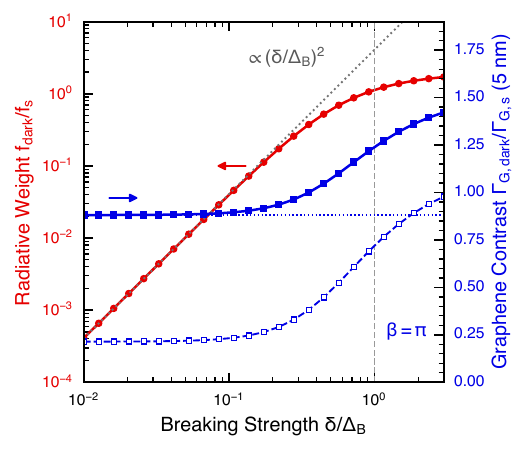}
\caption{Degenerate-doublet test of the $q=0$ darkness under generic
$C_3$ breaking $\delta$ protected by a $C_3$-preserving splitting
$\Delta_B$, from diagonalizing
$H_{\mathrm{eff}}=\tfrac{\Delta_B}{2}\sigma_z+\delta\,\sigma_x$ in the chiral
$p$-doublet basis.  Left axis: radiative weight of the dark-like eigenstate
relative to $s$, growing as $(\delta/\Delta_B)^2$ (dotted guide) in the
protected regime.  Right axis: its graphene contrast at $z=5$\,nm (filled squares,
solid: $\beta=0$; open squares, dashed: $\beta=\pi$), which stays at $\GG{p_+}/\GG{s}$
(dotted line, $\beta=0$) for $\delta/\Delta_B\ll1$ and rises toward the $p_+$/$p_-$
average once protection is lost (to the right of the dashed vertical line
$\delta=\Delta_B$).  All curves follow from an effective two-level
state-mixing model, which retains the unperturbed chiral decay matrices and
captures how state mixing lifts the darkness.  The texture-perturbation stress
tests of Fig.~S3 are complementary.}
\label{fig:c3break}
\end{figure}

\emph{Doublet-darkness protection (effective two-level model).}
Because $p_+$ and $d_-$ belong to \emph{degenerate} doublets, a generic
$C_3$-breaking perturbation $\delta$ (reconstruction, heterostrain, anisotropic
disorder) mixes each chiral pair at $O(1)$ however small $\delta$ is, unless a
$C_3$-preserving, complex-conjugation-breaking splitting $\Delta_B$ is applied
first (Sec.~\ref{sec:com}).  We test the doublet directly.  In the chiral basis
$\{p_+,p_-\}$ the perturbed manifold is
$H_{\mathrm{eff}}=\tfrac{\Delta_B}{2}\sigma_z+\delta\,\sigma_x$, which we
diagonalize (mixing angle $\tan2\vartheta=2\delta/\Delta_B$) and evaluate on the
dark-like eigenstate
$|\psi_{\rm d}\rangle=\cos\vartheta\,|p_+\rangle-\sin\vartheta\,|p_-\rangle$
(Fig.~\ref{fig:c3break}).  Its normal-incidence radiative weight is
$\sin^2\!\vartheta\,f_{p_-}$ and its graphene rate is
$\cos^2\!\vartheta\,\GG{p_+}+\sin^2\!\vartheta\,\GG{p_-}$ (both decay matrices are
diagonal in the chiral basis).  In the \emph{protected} regime $\delta\ll\Delta_B$ the
dark partner brightens only as $f_{\mathrm{dark}}/f_s\propto(\delta/\Delta_B)^2$
(a $10\%$ generic breaking gives ${\sim}4\times10^{-2}f_s$), and its graphene
contrast stays at $\GG{p_+}/\GG{s}\simeq0.88$ ($0.21$ for $\beta=\pi$).  As
$\delta/\Delta_B\to1$ the protection is lost and both the radiative weight and
the graphene contrast approach the $p_+$/$p_-$ average.  The two-bath selectivity thus survives generic
state-mixing imperfection \emph{provided} a $C_3$-preserving splitting protects
the chiral degeneracy.

\begin{figure*}[tbp]
\includegraphics[width=\textwidth]{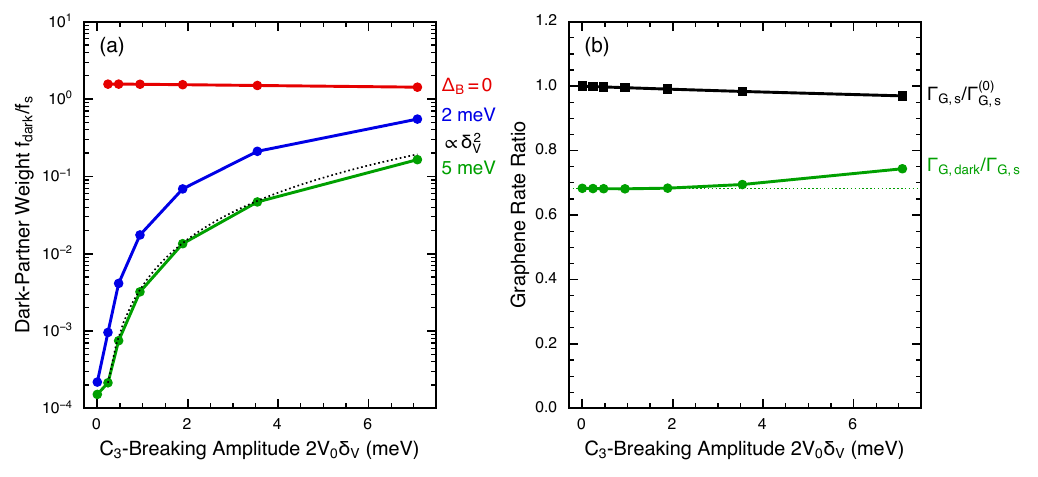}
\caption{Re-diagonalized $C_3$-breaking deformation of the moir\'e potential,
$V_{\mathrm M}\to V_{\mathrm M}+\delta_V\cdot2V_0\cos(\Gv_1\cdot\Rv+\phi_V)$, with the
trap re-solved and $\Pn{n}(\qv)$ rebuilt ($z=5$\,nm).  (a)~Photon weight of the
darker doublet member.  Without protection ($\Delta_B=0$) the degenerate pair
mixes at $O(1)$ for any $\delta_V>0$ and no member is dark (the point $\delta_V=0$ is
omitted for $\Delta_B=0$).  With a
$C_3$-preserving doublet splitting $\Delta_B=2$ and $5$\,meV [coupling
$-(\Delta_B/2)L_z$] the dark partner brightens quadratically (dotted guide).  (b)~The $s$ graphene rate relative to the
unperturbed value (black) and the dark partner's graphene contrast for
$\Delta_B=5$\,meV (green, as in (a); dotted: $\delta_V=0$).  The colors label
$\Delta_B$, not states.}
\label{fig:strain}
\end{figure*}

\emph{Re-diagonalized test.}  The texture stress tests of 
Supplemental Sec.~S1 deform $\Pn{n}(\qv)$ with the
moir\'e potential and orbitals held fixed.  Here the perturbation acts on the
potential itself and the orbitals are recomputed.  We add a representative
$C_3$-breaking deformation (a uniaxial heterostrain-like imbalance of one
first-shell harmonic),
$V_{\mathrm M}\to V_{\mathrm M}+\delta_V\cdot2V_0\cos(\Gv_1\cdot\Rv+\phi_V)$
with dimensionless amplitude $\delta_V$, and, as the $C_3$-preserving, complex-conjugation-breaking protection of
Sec.~\ref{sec:com}, the coupling $-(\Delta_B/2)L_z$ about the
trap center.  With this normalization $\Delta_B$ is the full doublet splitting
($L_z=\pm1$ partners shift by $\mp\Delta_B/2$), as in the two-level model
$H_{\mathrm{eff}}=\tfrac{\Delta_B}{2}\sigma_z+\delta\sigma_x$ above.  The computed splitting at $\delta_V=0$ is $1.98$ and
$4.95$\,meV for $\Delta_B=2$ and $5$\,meV.  The trap is re-solved by finite
differences in an open box ($0.34\,\aM$, $200^2$ points; the gauge-free
construction of Appendix~\ref{app:localization}), and $\Pn{n}(\qv)$, the photon
weights, and the graphene rates are rebuilt from the recomputed orbitals
[Fig.~\ref{fig:strain}].

\emph{$s$ channel.}  The $s$ orbital is nondegenerate and its graphene rate
[Fig.~\ref{fig:strain}(b)] changes by less than $3.0\%$ ($3.6\%$ for $\beta=\pi$) at $z=5$\,nm for
perturbation amplitudes $2V_0\delta_V\le7.1$\,meV.

\emph{Unprotected doublet.}  For $\Delta_B=0$ and any $\delta_V>0$ the solver
selects real ($p_x$/$p_y$-like) eigenstates that are $O(1)$ mixtures of the chiral
partners, and neither is dark, with $f/f_s\simeq1.4$--$1.6$ (the point
$\delta_V=0$ is omitted; see \emph{Numerical basis} below).  The bare splitting
that the perturbation induces is $\delta_{\rm split}=0.31\times2V_0\delta_V$
($=2\delta$ in the notation of the two-level model).  This is the disorder scale
that competes with the protection.

\emph{Protected doublet.}  With $\Delta_B=2$ and $5$\,meV the eigenstates stay
chiral and the darker member brightens [Fig.~\ref{fig:strain}(a)] as
$f_{\rm dark}/f_s\simeq f_{\rm box}+0.82\,(\delta_{\rm split}/\Delta_B)^2$, the coefficient being
$0.74$ and $0.91$ for the two protections ($f_{\rm box}\sim10^{-4}$ is the finite-box
floor).  Hence $f_{\rm dark}/f_s\le10^{-2}$ requires
$\Delta_B\ge9.1\,\delta_{\rm split}$, and $\le10^{-3}$ requires
$\Delta_B\ge29\,\delta_{\rm split}$.  Within this window the dark partner's
graphene contrast changes by about $1\%$ or less.  At $\delta_V=0$ it is $0.70$
($\Delta_B=2$\,meV) and $0.68$ ($5$\,meV) for $\beta=0$ and $0.19$ for $\beta=\pi$,
weakly dependent on $\Delta_B$ and $21$--$23\%$ ($10$--$12\%$) below the
Bloch-route values $0.88$ ($0.21$), an offset we attribute mainly to the finite
open box of this test (the single-well benchmark of
Appendix~\ref{app:localization} gives $0.816$--$0.873$ for $\beta=0$ depending on the
box).  Beyond the window the contrast rises toward the bright partner's (to
$0.94$, or $0.49$ for $\beta=\pi$, for $\Delta_B=2$\,meV at the largest
$\delta_V$).

\emph{Numerical basis.}  Inside an exactly degenerate subspace the eigensolver
returns an arbitrary basis.  At $\Delta_B=\delta_V=0$ this basis is a real mixture
of the $p$ pair, and individual oscillator strengths there are basis dependent.
Subspace-invariant quantities (the summed doublet oscillator strength and
the trace of the decay matrix) are well defined.  Thus, we report
single-state quantities only where $\delta_V>0$ or $\Delta_B>0$ lifts the
degeneracy, and chiral states only when $\Delta_B$ does.

\emph{The splitting $\Delta_B$.}  Moir\'e-trapped emitters show linewidths of
${\sim}100\,\mu$eV~\cite{Seyler2019}.  The splitting $\Delta_B$ of the $p_\pm$
doublet within one valley is a quantity to be measured.  The measured
valley-Zeeman $g$ factors describe a different splitting, and the neutral center
of mass has no orbital magneton.  Resolving the doublet line in field determines
$\delta_{\rm split}$ and $\Delta_B$ together, and the window
$\Delta_B\ge9.1\,\delta_{\rm split}$ then decides whether the dark partner is
protected.

\section{First-principles sign of the texture phase}
\label{sm:dft}

Symmetry restricts the relative phase of the two circular texture components to
$\beta=\arg(D_-/D_+)\in\{0,\pi\}$ (Sec.~\ref{sec:texture}), and the
published first-principles fits give only $(|D_+|,|D_-|)$~\cite{Wu2018}.
We determine the sign by a calculation of the aligned heterobilayer.

\emph{Structure and method.}  R-stacked WSe$_2$ (bottom, hole layer) and
MoSe$_2$ (top, electron layer) share the lattice constant $a_0=3.32$\,\AA{} and
a metal--metal distance $d_{\rm MM}=6.47$\,\AA~\cite{Wu2018}.  The chalcogen planes sit
$1.67$\,\AA{} above and below each metal plane, and the cell height is
$30$\,\AA.  We use the frame
$\bm a_1=a_0(1,0)$, $\bm a_2=a_0(\tfrac12,\tfrac{\sqrt3}{2})$, and
$+K=(2\bm b_1+\bm b_2)/3$, with $\bm b_1$, $\bm b_2$ the reciprocal vectors of
$\bm a_1$, $\bm a_2$.  The top layer is rigidly displaced by $\bm r_0$.  With
the bottom-layer chalcogen at $(\bm a_1+\bm a_2)/3$, the displacements
$\bm r_{0,n}=n(\bm a_1+\bm a_2)/3$ place Mo over W ($n=0$), over Se ($n=1$), and
over the empty site ($n=2$), as defined in Ref.~\onlinecite{Wu2018}.

The
calculations use the Vienna \emph{ab initio} simulation package
(VASP)~\cite{Kresse1996} with projector-augmented-wave (PAW)
potentials~\cite{Blochl1994,Kresse1999}, spin--orbit coupling in the
noncollinear formalism, the Perdew--Burke--Ernzerhof (PBE)~\cite{Perdew1996} or
local-density (LDA)~\cite{Ceperley1980,Perdew1981} functional, a $400$\,eV cutoff, and a $12\times12$ mesh, followed by a
non-self-consistent step at $\pm K$.  As a check we repeat the calculation
self-consistently with the Heyd--Scuseria--Ernzerhof hybrid functional
(HSE06)~\cite{Heyd2003,Krukau2006} on the $12\times12$ mesh, which
contains $\pm K$, with the exchange evaluated on a $6\times6$ subgrid.  At $+K$ we take the top valence band
(WSe$_2$ character) and the MoSe$_2$ conduction band with the spin of that valence band (the
spin-conserving, bright transition), identified by their layer and spin
projections, and evaluate the interband matrix element $\bm D(\bm r_0)=\langle
v|\hat{\bm p}|c\rangle$ both from the plane-wave coefficients alone and, including
the augmentation terms, from the optical matrix elements written by VASP.

\emph{Extraction.}  At fixed $\bm r_0$ a change of the Bloch phases multiplies
both circular components $c_\pm=\bm e_\pm^{*}\cdot\bm D$ by a common factor.
Thus, the ratio $c_-/c_+$ is gauge invariant.  No phase has to be transported
between different displacements.  The three-harmonic form of the texture
[Eq.~\eqref{eq:texture} at the monolayer scale] gives
$c_+=D_+\Sigma_0/3$ and $c_-=D_-\Sigma_1/3$ with
$\Sigma_0=\sum_n e^{-\ii\bm G_n\cdot\bm r_0}$,
$\Sigma_1=\sum_n e^{-\ii\bm G_n\cdot(\bm r_0-\bm r_{0,1})}$, and
$\bm G_n=\hat R^{\,n}K-K$ ($n=0,1,2$, with $\hat R$ the $120^\circ$ rotation), hence
$D_-/D_+=(c_-/c_+)(\Sigma_0/\Sigma_1)$ at any displacement where $\Sigma_0\Sigma_1\neq0$.

\emph{Checks.}  At the symmetric displacements the calculation reproduces the
selection rules of Ref.~\onlinecite{Wu2018}.  At $\bm r_{0,0}$ only $\sigma^+$
($|c_-/c_+|=4\times10^{-7}$), at $\bm r_{0,1}$ only $\sigma^-$
($|c_+/c_-|=1\times10^{-5}$), and at $\bm r_{0,2}$ both components vanish (below
$3\times10^{-6}$ of the $\bm r_{0,0}$ amplitude).  These zeros determine the
orientation, the valley, and the order of the matrix element.  Reversing any of
them gives $|D_-/D_+|$ near $5$, far above the ratio below one obtained here.

The
symmetric-point data also reproduce the parameters of Ref.~\onlinecite{Wu2018} used
here (PBE, $d_{\rm MM}=6.47$\,\AA, augmentation terms included).
These data give $|D_+|/D_0=0.215$ at $\bm r_{0,0}$ and $|D_-|/D_0=0.056$ at $\bm r_{0,1}$, with
$D_0$ the MoSe$_2$ intralayer matrix element at $\bm r_{0,0}$ ($0.22$ and $0.06$
in Ref.~\onlinecite{Wu2018}).  The interlayer gap at $\bm r_{0,0}$,
$\bm r_{0,1}$, $\bm r_{0,2}$ varies by $+13.4$, $-63.6$, and $+50.2$\,meV about
its mean, which gives $(V_0,\phi_V)=(11.2\,\mathrm{meV},78.5^\circ)$ compared
with $(11.8\,\mathrm{meV},79.5^\circ)$, with the deepest well at $R_1$ in both.  (In
the gap comparison we use a fixed $d_{\rm MM}$ and omit the registry dependence of
the exciton binding energy.)

\emph{Result.}  At the low-symmetry displacements the ratio $D_-/D_+$ is real and
positive in every case (Table~\ref{tab:dft}).  Its phase stays within
$1.6^\circ$ of zero for three functionals, three interlayer distances, and two
displacements, and the augmentation terms change its magnitude but not its sign.
Hence $\beta=0$.  The hybrid functional opens the gap at $K$ by about
$0.3$\,eV, and the ratio is unchanged.  A $500$\,eV cutoff
and an $18\times18$ mesh change $|D_-/D_+|$ by ${<}10^{-3}$ and the phase by
${<}0.1^\circ$.  Semicore W and Mo datasets change $|D_-/D_+|$ by $2\%$ ($7\%$ for
the plane-wave part alone), with the phase
remaining zero.

The real ratio at a general displacement is
an independent check of the symmetry argument.  Its magnitude there,
$0.32$--$0.35$, exceeds the symmetric-point ratio $|D_-|/|D_+|=0.26$ ($0.27$ in
Ref.~\onlinecite{Wu2018}).  The difference and the residual phase of
$1$--$2^\circ$ measure the accuracy of the three-harmonic form, and the sign is
independent of it.  The geometry is idealized (common lattice constant, fixed
chalcogen height, no relaxation).  However, the sign is unchanged across the tested
variations.  For the opposite sign the dark $p_+$ ratio
$\GG{p_+}/\GG{s}$ at $z=5$\,nm would drop from $0.88$ to $0.21$, its crossing
with $s$ would move below $2$\,nm, and $\alpha_{p_+}(10\,\mathrm{nm})$ would rise
from $5.5$ to $7.0$.  The other $\beta=\pi$ values are given alongside the
$\beta=0$ results in Appendices~\ref{sm:gauss} and~\ref{sm:strain} and throughout
the Supplemental Material.

\begin{table*}[tbp]
\caption{\label{tab:dft}%
First-principles texture ratio $D_-/D_+$ of aligned WSe$_2$/MoSe$_2$ at two
low-symmetry displacements $\bm r_0$ (in units of $\bm a_1,\bm a_2$), from the
plane-wave part of the matrix element and with the augmentation terms (PAW).
$d_{\rm MM}$: metal--metal distance.}
\begin{ruledtabular}
\begin{tabular}{lcccc}
functional, $d_{\rm MM}$ (\AA) & $\bm r_0$ & $|D_-/D_+|$ (plane wave) & $|D_-/D_+|$ (PAW) & $\arg(D_-/D_+)$ \\
\hline
PBE, $6.47$ & $(\tfrac16,\tfrac16)$ & $0.206$ & $0.335$ & $0.0^\circ$ \\
PBE, $6.47$ & $(0.30,0.10)$         & $0.205$ & $0.334$ & $1.1^\circ$--$1.6^\circ$ \\
PBE, $6.20$ & $(\tfrac16,\tfrac16)$ & $0.202$ & $0.328$ & $0.0^\circ$ \\
PBE, $6.80$ & $(\tfrac16,\tfrac16)$ & $0.216$ & $0.350$ & $0.0^\circ$ \\
LDA, $6.47$ & $(\tfrac16,\tfrac16)$ & $0.182$ & $0.319$ & $0.0^\circ$ \\
LDA, $6.47$ & $(0.30,0.10)$         & $0.180$ & $0.318$ & $0.9^\circ$--$1.4^\circ$ \\
HSE06, $6.47$ & $(\tfrac16,\tfrac16)$ & $0.205$ & $0.322$ & $0.0^\circ$ \\
HSE06, $6.47$ & $(0.30,0.10)$       & $0.202$ & $0.319$ & $1.0^\circ$--$1.4^\circ$ \\
\end{tabular}
\end{ruledtabular}
\end{table*}

\section{Model assumptions and robustness}
\label{app:robust}

\emph{Model assumptions.}
The theory rests on the single-channel COM model with the rigid,
leading-harmonic moir\'e potential and in-plane dipole texture at $\theta=1^\circ$.
Layer hybridization and field-driven brightening~\cite{Villafane2023,Polovnikov2024,Ouyang2026},
valley exchange (weak for interlayer excitons~\cite{Wu2018}), lattice
relaxation, and higher moir\'e harmonics are outside its scope.  We work in one
valley; the $-K$ valley follows by time reversal
($\sigma^+\leftrightarrow\sigma^-$, $p_+\leftrightarrow p_-$).  Charge transfer
and Dexter processes require $z\lesssim1$--$2$\,nm~\cite{Froehlicher2018,Lorchat2020,Moghe2026}
and are negligible for the hBN-spaced structures considered here
($z\geq2$\,nm).  The baseline graphene response is the clean $T=0$ RPA; the
causal finite-temperature broadening of Supplemental Sec.~S1 is a
sensitivity check.  The following tests quantify the sensitivity to these
assumptions.

\emph{Robustness.}
(i)~Replacing the projection-gauge Wannier orbitals by the gauge-free bound
states of a single trap changes the order-unity photon and graphene ratios by
at most $8\%$ (Appendix~\ref{app:localization}).  (ii)~A conservative relative-coordinate form
factor $F_{\mathrm{rel}}(q)=[1+(qa_X/2)^2]^{-3/2}$ with
$a_X=1.0$--$1.6$\,nm~\cite{Wu2018} moves $\GG{p_+}/\GG{s}$ at $z=5$\,nm from
$0.88$ to $0.80$--$0.85$ and $Q_s(5\,\mathrm{nm})$ from $23$ to $19$--$21$,
and every classification survives.  (iii)~Numerical refinement changes the
cancellation-sensitive observables by less than $10^{-4}$
(Appendix~\ref{app:numerics}, item~(vii)).

(iv)~The relative texture phase
enters every graphene rate through the interchannel interference
(Sec.~\ref{sec:kernels}).  Its sign $\beta=0$ is stable under all
first-principles variations of Appendix~\ref{sm:dft}, and the Supplemental
Material lists the changes for $\beta=\pi$.  The photon side
is independent of $\beta$.  (v)~$V_0$ and $M_X$ enter the trap states only through
their product.  A change of $\pm20\%$ in either shifts $\GG{n}/\GG{s}(5\,\mathrm{nm})$ by
${\le}21\%$ ($40\%$ for the small $d_+$ ratio), the crossings by $-17/{+}25\%$,
and $\alpha_n(10\,\mathrm{nm})$ by at most $0.1$, and the selection rules and the
inversion survive.

\emph{Stacking dependence.}
All results above are for the AA registry, whose deep well ($66.7$\,meV at
$R_1$) hosts the full $s$, $p_\pm$, $2s$, $d_\pm$ ladder.  AA and AB label the two
global stacking registrations of the WSe$_2$/MoSe$_2$ bilayer---the
near-$0^\circ$ and near-$60^\circ$ alignments, each with its own moir\'e
potential $(V_0,\phi_V)$---realized in \emph{separate} devices.  Within either device the local registry sweeps the
three high-symmetry sites across the $\aM\simeq19$\,nm cell, but the exciton
localizes in the deepest well and reports the registry at that minimum ($R_1$ for
AA, $R_0$ for AB).  The real-space integral behind Eq.~\eqref{eq:Pq} runs over the
whole cell, but the spatial support of $W_n(\Rv)$ is concentrated near the well
minimum.  Thus, the selectivity survives the averaging over the coexisting registries.

The AB registry provides a symmetry-level cross-check.  The trapping site and
the circular roles are interchanged ($s$ is $\sigma^+$, the dark partner is
$p_-$), and the $q=0$-dark partner remains graphene-bright at finite $q$
(Supplemental Sec.~S4).  The AB values take $\beta=0$, whereas the
first-principles calculation of Appendix~\ref{sm:dft} covers the aligned registry only.

\emph{Device stacks.}
A real device modifies the transfer kernel by a
$q$-dependent factor that does not cancel in the ratios.  Over eight
representative stacks, from image enhancement (air caps, a suspended stack) to
maximal screening (metallic gates), the distance-driven
inversion and $\alpha_{p_+}>\alpha_s$ survive, and the order of a
state relative to $s$ flips only near its crossing.  The crossings and
exponents shift at the $10$--$40\%$ level (Supplemental Sec.~S2).  The radiative side of the layered problem (the full Green tensor) is outside
the present model.

\bibliography{refs}

\end{document}